\documentclass[12pt]{article}
\usepackage[margin=2.5cm]{geometry}
\usepackage[utf8]{inputenc}
\usepackage{amsmath}
\usepackage{amssymb}
\usepackage{hyperref}
\usepackage{graphicx}
\usepackage{colortbl}
\usepackage{multirow}
\usepackage{array}
\usepackage{lineno}

\usepackage[backend=biber,style=nature, natbib=true]{biblatex}
\usepackage{graphicx}

\usepackage{url}

\usepackage{algorithm}
\usepackage{algpseudocode}

\usepackage{tikz}
\usepackage[export]{adjustbox}
\usepackage{caption}
\usepackage[skip=1ex, belowskip=2ex]{subcaption}
\usetikzlibrary{arrows.meta}

\usepackage{pdflscape}
\usepackage{afterpage}

\usepackage{bbold}

\usepackage{array}
\newcolumntype{L}[1]{>{\raggedright\let\newline\\\arraybackslash\hspace{0pt}}m{#1}}
\newcolumntype{C}[1]{>{\centering\let\newline\\\arraybackslash\hspace{0pt}}m{#1}}
\newcolumntype{R}[1]{>{\raggedleft\let\newline\\\arraybackslash\hspace{0pt}}m{#1}}

\DeclareMathOperator*{\argmin}{arg\,min}

\newcommand{\LL}{\mathrm{NLL}}

\newcommand{\M}{\mathcal{M}}
\newcommand{\Mopt}{\mathcal{M}_{I^*}}
\newcommand{\Mcand}{\mathcal{M}_{I}}
\newcommand{\Msub}{\mathcal{M}_{\Tilde{I}}}
\newcommand{\Msup}{\mathcal{M}_\mathbb{1}}
\newcommand{\MoptCurr}{\mathcal{M}_{I^+}}
\newcommand{\Mnext}{\mathcal{M}_{\text{next}}}

\newcommand{\Mcal}{\mathfrak{M}_{\text{calibrated}}}
\newcommand{\Mex}{\mathfrak{M}_{\text{exclusions}}}
\newcommand{\Mexable}{\mathfrak{M}_{\text{excludable}}}
\newcommand{\Munseen}{\mathfrak{M}_{\text{unseen}}}
\newcommand{\Mall}{\mathfrak{M}_{\text{all}}}

\newcommand{\Crit}{\mathcal{C}}
\newcommand{\D}{\mathcal{D}}
\newcommand{\Fit}{\mathcal{F}}
\newcommand{\Pen}{\mathcal{P}}
\newcommand{\Bound}{\mathcal{B}}
\newcommand{\nBound}{\bar{\mathcal{B}}}

\newcommand{\bound}{bounding}
\newcommand{\BOUND}{Bounding}

\newcommand{\Aterm}{\mathcal{T}}
\newcommand{\Aplaus}{\mathcal{P}}

\newcommand{\Cp}{\mathit{C_p}}

\newcommand{\AIC}{\text{AIC}}
\newcommand{\AICc}{\text{AIC}_\text{c}}
\newcommand{\BIC}{\text{BIC}}

\newcommand{\J}[1][{}]{\mathcal{J}
\ifthenelse { \equal{#1}{} }  %if short caption not specified, use long caption (no slant)
    { }   % if #1 == blank
    { (m_{#1}) }   % else (not blank)
}

\providecommand{\keywords}[1]
{
  \small	
  \textbf{\textit{Keywords---}} #1
}

\title{Abstract, keywords and references template}
\author{Author Surname$^{1}$, Someone Else$^{2}$  \\
        \small $^{1}$University A \\
        \small $^{2}$University B \\
}

\usepackage{xcolor}

\newcommand{\supp}{Supplementary information}

\usepackage{orcidlink}

\title{Scalable branch-and-bound model selection with non-monotonic criteria including AIC, BIC and Mallows's $\Cp$}
\author{
Jakob Vanhoefer\,$^{\text{1}}$\, \orcidlink{0000-0002-3451-1701},
Antonia Körner\,$^{\text{1}}$\, \orcidlink{0009-0005-9493-9936
}, 
Domagoj Doresic\,$^{\text{1}}$\, \orcidlink{0009-0007-1820-7375
},\\
Jan Hasenauer\,$^{\text{1}, \ast}$\, \orcidlink{0000-0002-4935-3312
},
Dilan Pathirana\,$^{\text{1}, \ast, L}$\, \orcidlink{0000-0001-7000-2659}
}
\date{}

\begin{document}

\maketitle
{\small
$^{\text{1}}$ Life and Medical Sciences (LIMES) Institute and Bonn Center for Mathematical Life Sciences, University of Bonn, Bonn, Germany\\
$^\ast$ Corresponding authors: jan.hasenauer@uni-bonn.de and dilan.pathirana@uni-bonn.de\\
$^L$ Lead contact.
}

%%%%

%\linenumbers

\begin{abstract}
Model selection is a pivotal process in the quantitative sciences, where researchers must navigate between numerous candidate models of varying complexity. Traditional information criteria, such as the corrected Akaike Information Criterion (AICc), Bayesian Information Criterion (BIC), and Mallows's $\Cp$, are valuable tools for identifying optimal models. However, the exponential increase in candidate models with each additional model parameter renders the evaluation of these criteria for all models -- a strategy known as exhaustive, or brute-force, searches -- computationally prohibitive. Consequently, heuristic approaches like stepwise regression are commonly employed, albeit without guarantees of finding the globally-optimal model.

In this study, we challenge the prevailing notion that non-monotonicity in information criteria precludes bounds on the search space. We introduce a simple but novel bound that enables the development of branch-and-bound algorithms tailored for these non-monotonic functions. We demonstrate that our approach guarantees identification of the optimal model(s) across diverse model classes, sizes, and applications, often with orders of magnitude computational speedups. For instance, in one previously-published model selection task involving $2^{32}$ (approximately 4 billion) candidate models, our method achieves a computational speedup exceeding 6,000. These findings have broad implications for the scalability and effectiveness of model selection in complex scientific domains.

\end{abstract}

\keywords{Model selection, branch-and-bound, information criteria, Akaike information criterion (AIC), Bayesian information criterion (BIC), Mallows's $\Cp$}

\section*{Background}

Mathematical models are of fundamental importance for quantitative science in general, and the physical and life sciences in particular \citep{Mesterton2011, Kitano2002, Burnham2004}. These models formalise assumptions on the data-generating process, e.g., a chemical reaction network, a cellular signalling pathway, competition in an ecological system, or a pandemic. 
However, it's often unclear whether to include certain system components in the model, and how exactly a component should be represented mathematically, leading to different candidate models describing different hypotheses \citep{Saltelli2019, Bodner2021, Villaverde2022}.

Model selection is the task of finding optimal model(s) from a set of candidate models, given some data. The task has been extensively studied and has attracted significant attention in the scientific literature. Early works by \cite{Jeffreys1961}, \cite{Akaike1973} and \cite{Schwarz1978} were followed by many extensions, reviews and practitioners guides \citep{Burnham2004, Ding2018}. The set of candidate models is determined by design choices on mechanisms or causalities in the data generation process. Commonly, every individual choice translates into a term being present or absent in the mathematical model. As a result, the set of candidate models (i.e., model space) grows by a factor of two for each choice, i.e., exponentially with the number of choices \citep{Kirk2013}.

A common approach to compare models is to use a criterion such as the Akaike information criterion (AIC) \citep{Akaike1973}, corrected AIC ($\AICc$) \citep{Hurvich1989}, Bayesian information criterion (BIC) \citep{Schwarz1978} or Mallows's $C_p$ \citep{Mallows1973}. These information criteria balance the goodness of fit of a model with its complexity. Given such criteria, the search for the best model can be formulated as a mixed-integer nonlinear optimization problem for discrete design variables and continuous parameters \citep{Rodriguez2013}.

There are two classes of methods for solving such optimization problems: global and local methods \citep{Calcagno2010}. The global brute-force method guarantees that the optimal model is found, by computing the criterion (e.g., AIC) of each model in the model space. This requires reliable calibration of individual models, which is the most computationally costly step. Due to the exponential complexity of the model space, global approaches quickly become infeasible as the number of choices grows. In contrast, iterative, local methods are faster as they only calibrate a subset of the candidate models, but do not guarantee that optimal model(s) are found \citep{Grafen2002}. Currently, there is no method that is fast and guarantees that the optimal model(s) are found.

In this study, we propose a simple approach that is similar to branch-and-bound wrapper methods that have previously been used for feature selection (a specific case of model selection) \citep{furnivalRegressionsLeapsBounds1974,narendraBranchBoundAlgorithm1977}. However, in contrast to the available algorithms \citep{kohaviWrappersFeatureSubset1997}, the proposed approach is applicable to criteria such as AIC, $\AICc$, BIC, and Mallows's $\Cp$, which are non-monotonic with respect to the number of parameters, i.e., they are non-monotonic because adding parameters to a model may worsen the criteria. The proposed approach facilitates the use of branch-and-bound methods for popular but non-monotonic model selection criteria.

We use the criterion value of a previously calibrated model to derive a bound on the criteria of its uncalibrated submodels (i.e., nested models). We use this to develop branch-and-bound methods that often avoid the calibration of most models, for both local and global search strategies. We show that this approach can lead to speedups of $>6\,000$, compared to the brute-force exhaustive search, in realistic applications from different fields and with different model classes and sizes.

\section*{Results}

\subsection*{A bound on non-monotonic information criteria for uncalibrated submodels provides an \textit{a priori} condition for their exclusion}
The model space of any model selection problem can be defined as a subset of all potential submodels of the superset model, which contains all model design variables (parameters). This encompasses various linear and nonlinear regression and classification models. We use an indicator vector $I \in \{0,1\}^n$ to indicate the presence or absence of the $n$ possible design variables in a model $\Mcand$. Hence, the model space has size $\leq 2^n$, which is the exponential complexity that renders global methods infeasible for large $n$. We note that there can be additional estimated parameters that are included in all models, but these are not design variables and are therefore excluded from $I$, meaning that the $n$-dimensional $I$ can be lower dimensional than the set of all model parameters $\theta_{\Mcand}$. 

We refer to model $\Msub$ as a submodel of $\M_I$, if for all $i \in \{1, ..., n\}$ it holds that $\Tilde{I}_i \leq I_i$. Accordingly, a submodel $\Msub$ of $\M_I$ contains a subset of the parameters that appear in the model $\M_I$. If $\Msub$ is a submodel of $\M_I$, it holds that $\M_I$ is a supermodel if $\Msub$. The superset model is $\Msup$, with $\mathbb{1}$ denoting the indicator vector containing only ones. Every candidate model is a submodel of the superset, $\Mcand \leq \Msup$.

Information criteria $\Crit: \M, \D \rightarrow \mathbb{R}$ that are used to compare candidate models given a dataset $\D$, such as the AIC, $\AICc$, BIC, or Mallows's $\Cp$, share the structure:
\begin{align}
    \Crit(\Mcand, \D) = -s \Fit(\Mcand, \D) + \Pen(|\Mcand|, |\D|),
    \label{eq:criterion}
\end{align}
with constant scalar $s$, goodness-of-fit measure $\Fit: \Mcand, \D \rightarrow \mathbb{R}$ and penalty term $\Pen: |\Mcand|, |\D| \rightarrow \mathbb{R}$. The evaluation of the goodness-of-fit measure $\Fit(\Mcand, \D)$ requires the calibration of $\Mcand$ to $\D$, which can be computational demanding. In contrast, the penalty term $\Pen$ can be computed without calibration and depends merely on the number of estimated model parameters ($|\Mcand|:=$ number of estimated values in $\theta_{\Mcand}$), and the number of data points, $|\D|$.

The optimal model, $\Mopt=\argmin_{\M_I} \Crit(\M_I, \D)$, is the model for which the value of the chosen information criterion is lowest. In the search for $\Mopt$, local and global search methods explore the model space and store the best model seen so far $\MoptCurr$. A candidate model $\Mcand$ is calibrated to compute $\Crit$. However, if it is known prior to calibration that $\Crit(\Mcand, \D) > \Crit(\MoptCurr, \D)$, i.e., if the candidate model is worse than the best seen so far, then the calibration of $\Mcand$ can be skipped. This reduces the required computational resources.

Here, we propose a simple \textit{a priori} bound for the information criteria to reduce the number of model calibrations. This bound builds on the fact that the goodness-of-fit for a model $\Msub$, $\Fit(\Msub, \D)$, is at most as good as the goodness-of-fit for a model $\Mcand$, $\Fit(\Mcand, \D)$, if $\Msub$ is a submodel of $\Mcand$. This is true as $\Msub$ has a reduced set of degrees of freedom compared to $\Mcand$. From this bound on the goodness-of-fit of $\Msub$, a bound on the values of the information criteria directly follows, 
\begin{align*}
\Crit(\Msub, \D) &\geq \Bound_\Crit(\Msub, \Mcand) := -s \Fit(\Mcand, \D) + \Pen(|\Msub|, |\D|).
\end{align*}
If the condition
\begin{equation}
    \Bound_\Crit(\Msub, \Mcand) > \Crit(\MoptCurr, \D)
    \label{eq:condition}
\end{equation}
is met, then $\Msub$ can be excluded from the set of potentially optimal models without calibration. Importantly, the bound $\Bound_\Crit(\Msub, \Mcand)$ and condition can be evaluated for all uncalibrated $\left\{\Msub:\Msub \leq \Mcand\right\}$ after calibration of $\Mcand$, which can result in the exclusion of many uncalibrated models.
$\Msub$ is excluded if the condition is satisfied with any of its supermodels. Hence, if multiple supermodels have already been calibrated, then it is sufficient to check the condition with only the supermodel that maximises the bound.
If, in addition to the optimal model, all plausible models within some criterion threshold are desired, then the condition can be adjusted to achieve this, as described in the \supp{}.

Note that in this case of non-monotonic criteria, the condition \eqref{eq:condition} must be evaluated for every submodel size ($|\Mcand|$). This differs from branch-and-bound methods in the literature for monotonic criteria, where all submodels can be excluded at once.
Furthermore, $\MoptCurr$ changes whenever a model is found that improves on the ``best model seen so far''. This means additional uncalibrated models can be excluded by re-evaluating the condition whenever $\MoptCurr$ changes. Note also that the submodels of $\MoptCurr$ cannot be excluded, unless they are also submodels of another calibrated model.

\subsection*{The bound on the criterion can be simplified into a bound on model size}

The concept outlined in the previous section works for a broad range of information criteria. However, it can be further simplified for specific criteria to improve intuition and computation. Here, we provide the results for the AIC, and results for other information criteria are provided in the \supp{}.

The AIC is defined as $\AIC(\Mcand, \D) = 2\LL(\Mcand, \D) + 2|\Mcand|$, with $\LL(\Mcand, \D)$ denoting the negative log-likelihood evaluated at the maximum likelihood estimate. In terms of \eqref{eq:criterion}, the AIC is obtained for $s=2$, $-\Fit(\M, \D)=\LL(\M, \D)$, and $\Pen(\Mcand, \D):=2|\Mcand|$. Accordingly, a lower bound for the information criteria of an uncalibrated submodel $\Msub (\leq \Mcand)$ of the calibrated model $\Mcand$ is
\begin{equation}
    \AIC(\Msub, \D) \geq \Bound_\AIC(\Msub, \Mcand) = 2\LL(\Mcand, \D) + 2|\Msub|.
    \label{eq:bound_aic}
\end{equation}
To check whether the calibration of $\Msub$ can be skipped, the lower bound $\Bound_\AIC(\Msub, \Mcand)$ is compared to the AIC of the currently best model, $\AIC(\MoptCurr, \D)$. The rearrangement of the resulting inequality condition (cf. \eqref{eq:condition}) yields
\begin{align}
|\Msub| &> \frac{1}{2}\AIC(\MoptCurr,\D) - \LL(\Mcand, \D), \\
&= \frac{1}{2}\left(\AIC(\MoptCurr,\D) - \AIC(\Mcand,\D)\right) + |\Mcand|,
\label{eq:aic_cutoff}
\end{align}
\sloppy which implies that all uncalibrated submodels $\left\{\Msub:\Msub \leq \Mcand\right\}$ with more than $\nBound_\AIC:=\frac{1}{2}\left(\AIC(\MoptCurr,\D) - \AIC(\Mcand,\D)\right) + |\Mcand|$ parameters can be excluded. That is, the bound on the criterion $\Bound_\AIC(\Msub, \Mcand)$ can be reformulated into a bound on the size $|\Msub|$, $\nBound_\AIC$.

The simplified expression \eqref{eq:aic_cutoff} reveals that if the criterion of a calibrated model is much worse than the currently best model (i.e., if $\AIC(\MoptCurr, \D) - \AIC(\Mcand, \D)$ is large), many submodels of $\Mcand$ can be excluded. The larger submodels are excluded by the condition, meaning that parsimonious models are more likely to be calibrated. This is beneficial as calibration for these models is usually computationally less demanding. Additionally, if there are multiple supermodels ($\Mcand$), the worst will maximise $\AIC(\MoptCurr, \D) - \AIC(\Mcand, \D)$ and hence generally permit the exclusion of the most submodels (assuming each $\Mcand$ has the same number of submodels at each model size).

The condition expressions also provide interesting insights regarding model search strategies. As the exclusion of models depends critically on the value of the information criterion of the currently best model, model selection strategies that quickly improve this value might be favourable over alternative strategies. Accordingly, advanced local search strategies might be a good starting point and perform superior to random search, or forward and backward selection schemes.
In the next section, we provide further details on these approaches.

\subsection*{Development of local and global search methods using the proposed bound}

The results of the previous section suggest that established local and global search algorithms for model selection can benefit from the simple bound. To assess this in the subsequent sections, we implemented versions of three established algorithms that additionally exploit the bounds: two local search algorithms (backward selection and Flexible and dynamic Algorithm for Model Selection (FAMoS)) and one global search algorithm (brute-force exhaustive search). Therefore, we introduce the notion of an exclusion set, 
\begin{equation}
\{\Msub: |\Msub| \leq |\Mcand| \wedge \Bound_\Crit(\Msub, \Mcand) > \Crit(\MoptCurr, \D)\},
\label{eq:exclusions}
\end{equation}
which are the submodels of a calibrated model $\Mcand$ that can be excluded by the bound and condition.
Here, we briefly outline the algorithms, with high-level descriptions of the implementations. Full descriptions are provided in \textit{Methods}.

(Algorithm~1) The \textbf{\bound{} backward} selection method is a refinement of the popular backward selection method. Similar to backward selection, the proposed method attempts to find the optimal model by sequentially removing individual parameters from the currently best model. However, instead of directly performing the calibration for a model candidate, it checks if the model is contained in the exclusion set of any of the already calibrated models. If this is the case, the model is skipped. As in most implementations of backward selection, our \bound{} backward selection method first calibrates the uncalibrated model with the largest number of parameters. As this model also has the largest number of submodels, this strategy increases the possible number of exclusions.

(Algorithm~2) The \textbf{\bound{} FAMoS} method is a refinement of the Flexible and dynamic Algorithm for Model Selection (FAMoS) method. FAMoS is a local method that implements various meta-heuristics, for example, the possibility to jump to the most distant models in the model space (see \citep{Gabel2019} for a full description). FAMoS was previously found to outperform other standard local searches by finding better models \citep{Gabel2019}. Accordingly, the exclusion sets can be substantially larger after the same number of model calibrations. The \bound{} FAMoS local method is a re-implementation of FAMoS that skips models in any exclusion set.

(Algorithm~3) The \textbf{\bound{} exhaustive} method is a massive refinement of standard brute-force exhaustive search. Similar to exhaustive search, the proposed method explores the complete model space. However, instead of calibrating all model candidates, it uses a sequence of local searches that exploit the proposed bounds. Specifically, the method initiates an efficient local search at some point in model space. This search will ultimately terminate because (a) the full model space has been explored -- meaning that each model has either been calibrated or excluded -- or (b) no further improvement is achieved. Given (a), the exhaustive search is completed, whereas given (b), another local search -- which can exploit the information from previous searches -- is initiated with a model candidate which was neither calibrated nor excluded in previous local searches. The initiation of local searches is continued until all models are either calibrated or excluded. Thus, this method covers the entire model space. The local search strategies can be selected by the user, for example, some combination of the (\bound{}) backward or FAMoS methods.

Assuming that the model calibrations converge to the maximum likelihood estimate, it holds that (i) standard and \bound{} backward selection provide the same result, (ii) standard and \bound{} FAMoS provide the same result, and (iii) standard and \bound{} exhaustive search provide the same result (i.e. the optimal model). However, it is important to note that the number of model calibrations required by the \bound{} variants might be reduced due to the exclusion sets.

\subsection*{Illustration of the backward vs. \bound{} backward selection methods}

\begin{figure}[!t]
\begin{tabular}[t]{@{}l@{}l@{}l@{}}
\begin{tabular}[t]{l}
\noindent\textbf{(a)} Model space.\\
    \includegraphics[width=0.25\textwidth, valign=m]{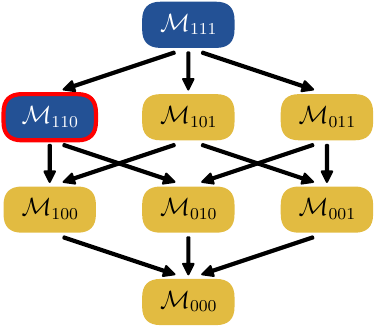}
\end{tabular}
&\begin{tabular}[t]{l}
    Legend\vspace{0.2cm}\\
    {\small
\begin{tabular}[t]{cc}
    \includegraphics[width=0.06\textwidth, valign=m]{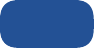}&Good AIC ($<-15$)\\
    \includegraphics[width=0.06\textwidth, valign=m]{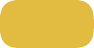}&Bad AIC ($>45$)\\
    \includegraphics[width=0.06\textwidth, valign=m]{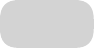}&Uncalibrated\\
    \includegraphics[width=0.06\textwidth, valign=m]{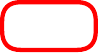}&Best seen so far ($\MoptCurr$)\\
    \includegraphics[width=0.06\textwidth, valign=m]{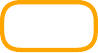}&Excluded by condition
\end{tabular}
}\\
Parent model \raisebox{1pt}{\tikz\draw[ultra thick,-{Latex[round,length=5pt, width=5pt]}] (0,0) -- ++ (0.8,0);} Child model
\end{tabular}
&\begin{tabular}[t]{l}
    \textbf{(b)} Best fit ($\M_{110}$) to data.\\
    \includegraphics[width=0.3\textwidth, valign=m]{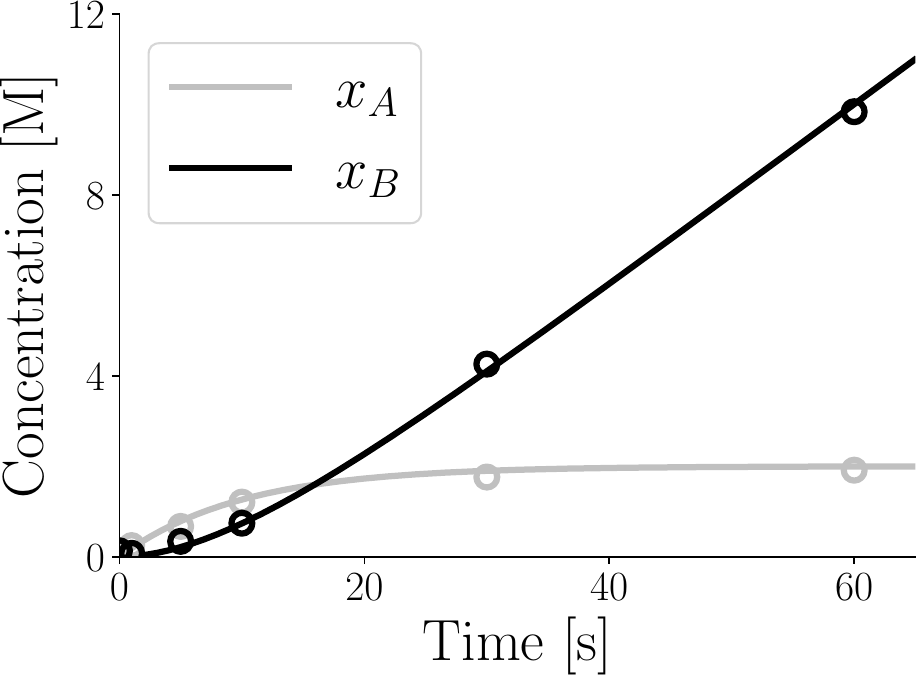}
    \end{tabular}
\end{tabular}\\

\noindent{\textbf{(c)} Iterations of backward model selection.\par}
{\centering
    \begin{subfigure}{0.25\textwidth}
    \includegraphics[width=\hsize, valign=m]{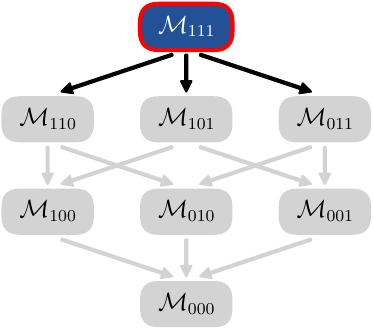}
    \caption*{Iteration 1}
    \end{subfigure}
\tikz[baseline=-5\baselineskip]\draw[gray, ultra thick,->] (0,0) -- ++ (1,0);
    \begin{subfigure}{0.25\textwidth}
    \includegraphics[width=\hsize, valign=m]{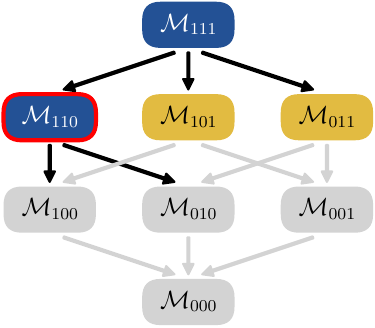}
    \caption*{Iteration 2}
\end{subfigure}
\tikz[baseline=-5\baselineskip]\draw[gray, ultra thick,->] (0,0) -- ++ (1,0);
    \begin{subfigure}{0.25\textwidth}
    \includegraphics[width=\hsize, valign=m]{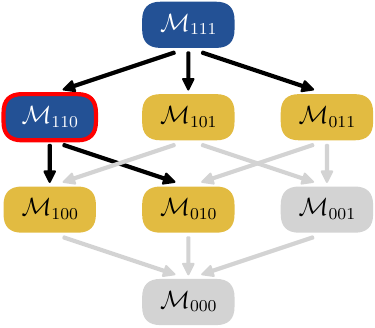}
    \caption*{Iteration 3}
\end{subfigure}
\par}

\noindent{\textbf{(d)} Iterations of \bound{} backward model selection.\par}
{\centering
    \begin{subfigure}{0.25\textwidth}
    \includegraphics[width=\hsize, valign=m]{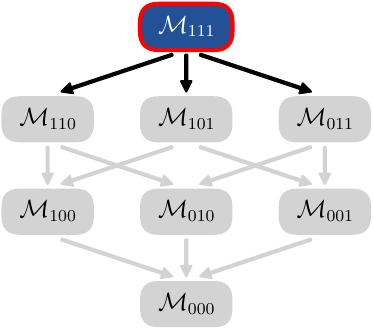}
    \caption*{Iteration 1\\\,}
    \end{subfigure}
\tikz[baseline=-6\baselineskip]\draw[gray, ultra thick,->] (0,0) -- ++ (1,0);
    \begin{subfigure}{0.25\textwidth}
    \includegraphics[width=\hsize, valign=m]{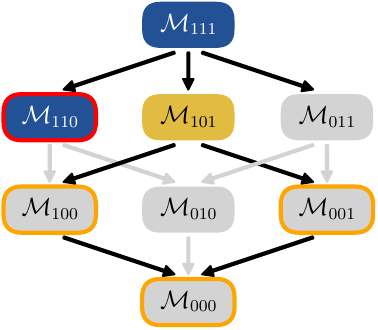}
    \caption*{\centering Iteration 2 (partial)\\With $\M_{101}$ exclusions}
\end{subfigure}
\tikz[baseline=-6\baselineskip]\draw[gray, ultra thick,->] (0,0) -- ++ (1,0);
    \begin{subfigure}{0.25\textwidth}
    \includegraphics[width=\hsize, valign=m]{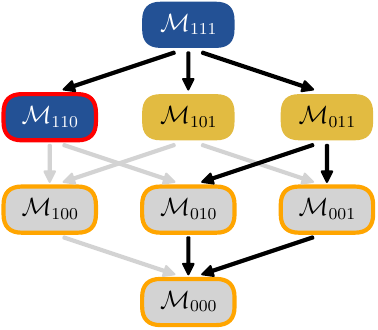}
    \caption*{\centering Iteration 2\\With $\M_{011}$ exclusions}
\end{subfigure}
\par}

    \caption{\textbf{Backward and \bound{} backward selections, with the small-scale model selection problem} \textbf{(a)} The model space, colored by the AIC of each model. Arrows (\protect\tikz\protect\draw[ultra thick,-{Latex[round,length=5pt, width=5pt]}] (0,0) -- ++ (0.8,0);) indicate submodels with 1 fewer parameter. \textbf{(b)} Synthetic data (open circles) and simulated state trajectories (lines) from the calibrated ``correct'' model $\M_{110}$. \textbf{(c, d)} Iterations of the backward and \bound{} backward selections.}
    \label{fig:small_scale_problem}
\end{figure}

To illustrate the usefulness of the bound and the properties of the proposed search methods, we consider a small-scale model selection problem. The physical context is a system with conversion and reaction processes involving two chemical species, $x_A$ and $x_B$. There are three design variables for the hypotheses of: ($\theta_1$) constant inflow of $x_A$, ($\theta_2$) conversion of $x_A$ into $x_B$, and ($\theta_3$) degradation of $x_A$ and $x_B$ upon their interaction. The model space is given by a collection of ordinary differential equation (ODE) models that each contain two state variables and up to three of the design variables in the parameter vector $\theta$. The superset model $\Msup$ is 
\begin{align}
&\xrightarrow{\theta_1} x_A \\
x_A &\xrightarrow{\theta_2} x_B \\
x_A + x_B &\xrightarrow{\theta_3} \\
\frac{dx_A}{dt} &= \theta_1 - (\theta_2+\theta_3 x_B) x_A,  && x_A(0) = 0, \\
\frac{dx_B}{dt} &= (\theta_2+\theta_3 x_B) x_A,  && x_B(0) = 0.
\end{align}
Given three design variables in $\theta$, the model space is $\left\{\Mcand : I \in \{0, 1\}^3\right\}$, for a total of $2^3=8$ unique models. We use a binary string to represent the indicator vector, where 1 indicates that the corresponding parameter is estimated, whereas 0 indicates that the parameter is removed from the model (in this case, set to zero). For example, $\Msup$ is $\M_{111}$ with $I = (1,1,1)$, indicating that each of $\theta_1$, $\theta_2$, and $\theta_3$ are estimated, respectively. $\M_{101}$ estimates $\theta_1$ and $\theta_3$, and sets $\theta_2=0$, i.e. $\M_{101}$ has $n_{101}=2$ parameters. All models arranged by complexity ($|\Mcand|$) are visualised in Figure~\ref{fig:small_scale_problem}a. Synthetic data of the full state at $t= \{0, 1, 5, 10, 30, 60 \}$ were generated with $\theta=(0.2,0.1,0)$ and additive Gaussian noise (standard deviation $\sigma = 0.15$)(Figure~\ref{fig:small_scale_problem}b). The measurement noise parameter $\sigma$ is estimated in all models and is therefore not a design variable.

We performed model selection to identify the model that generated the synthetic data. We used the AIC and compared the backward and \bound{} backward selections, which are both local methods. All employed model selection methods are specified in the \supp{}. Briefly, backward selections start at the superset model $\M_{111}$ and remove one parameter at a time, according to the criterion.

The backward selection terminated after calibrating 6 models, and identified the correct model (Figure~\ref{fig:small_scale_problem}c). The \bound{} backward selection achieved the same result after calibrating 4 models (Figure~\ref{fig:small_scale_problem}d). All submodels of $\M_{101}$ and $\M_{011}$ were excluded by the condition \eqref{eq:condition}, which is why the 2 additional models calibrated by the backward selection were skipped in the \bound{} backward selection. Interestingly, the \bound{} backward selection actually guaranteed that the correct model was found, because all uncalibrated models were excluded by the condition \eqref{eq:condition}. The backward selection would need to additionally calibrate $\M_{000}$ and $\M_{001}$ (i.e., 4 additional models in total, compared to the \bound{} backward selection) to make the same guarantee.

This illustrates that the \bound{} backward selection is able to exclude regions of the model space, which not only accelerates the search, but can in some cases guarantee identification of the optimal model.

\subsection*{\BOUND{} methods outperform standard methods on a large-scale synthetic problem}
\label{sec:famos_problem}

\begin{figure}[!ht]

\begin{minipage}[t]{.6\textwidth}
\textbf{(a)}\\\hphantom{\textbf{(C)}}
\begin{tabular}[t]{ll}
\begin{tabular}[t]{l}
    {\small Superset model.}\\
    \includegraphics[width=0.3\linewidth, valign=m]{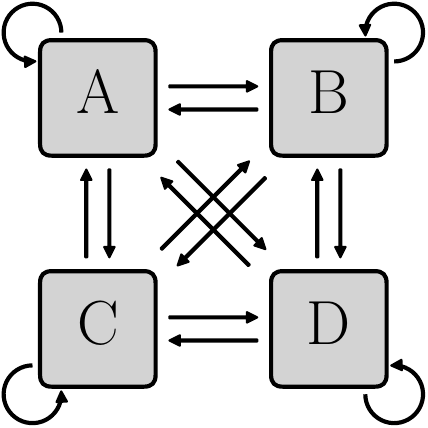}
\end{tabular}
&\begin{tabular}[t]{l}
    {\small True model.}\\
    \includegraphics[width=0.3\linewidth, valign=m]{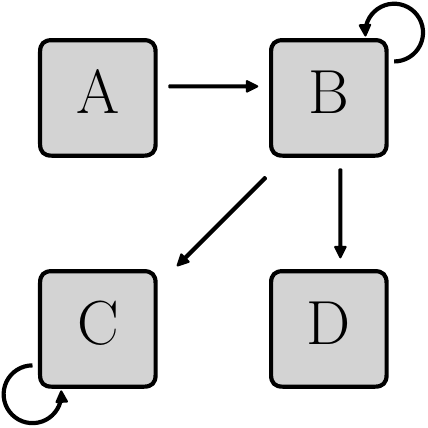}
    \end{tabular}
\end{tabular}
\end{minipage}
\begin{minipage}[t]{.4\textwidth}
    \textbf{(d)}\\\hphantom{\textbf{(C)}}
    \includegraphics[scale=0.5, valign=t]{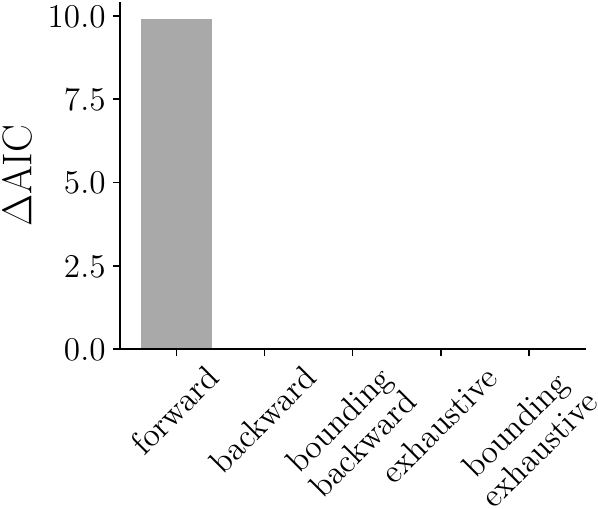}
\end{minipage}

%\vspace{0.5cm}

\begin{minipage}[t]{.6\textwidth}
\begin{tabular}[t]{l}
\textbf{(b)}\\\hphantom{\textbf{(C)}}
    \includegraphics[scale=0.5, valign=m]{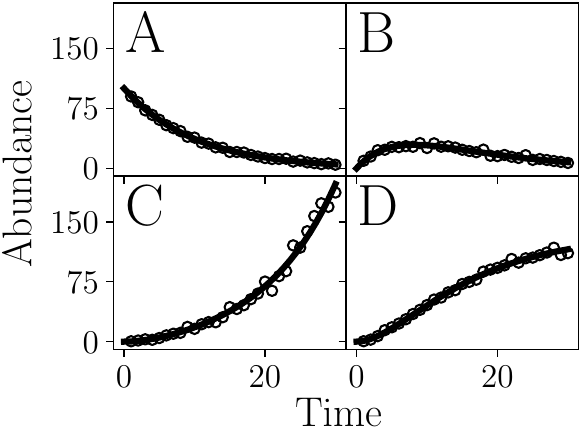}
\end{tabular}
\end{minipage}
\begin{minipage}[t]{.4\textwidth}
    \textbf{(e)}\\\hphantom{\textbf{(C)}}   
    \includegraphics[scale=0.5, valign=t]{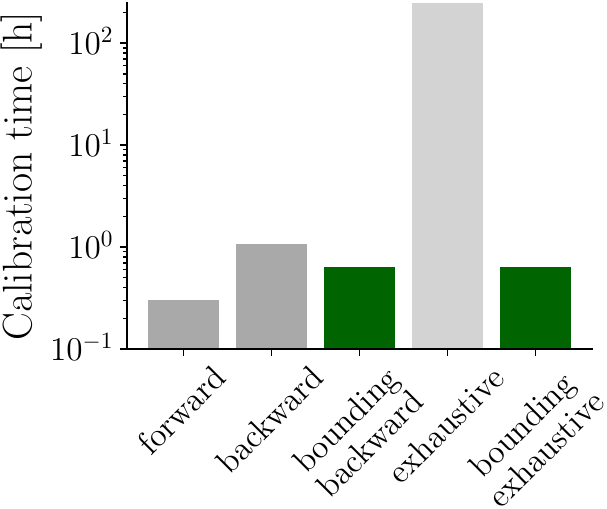}
\end{minipage}

\begin{minipage}[t]{.6\textwidth}
\begin{tabular}[t]{l}
    \textbf{(c)}\\\hphantom{\textbf{(C)}}
    \includegraphics[scale=0.5, valign=t]{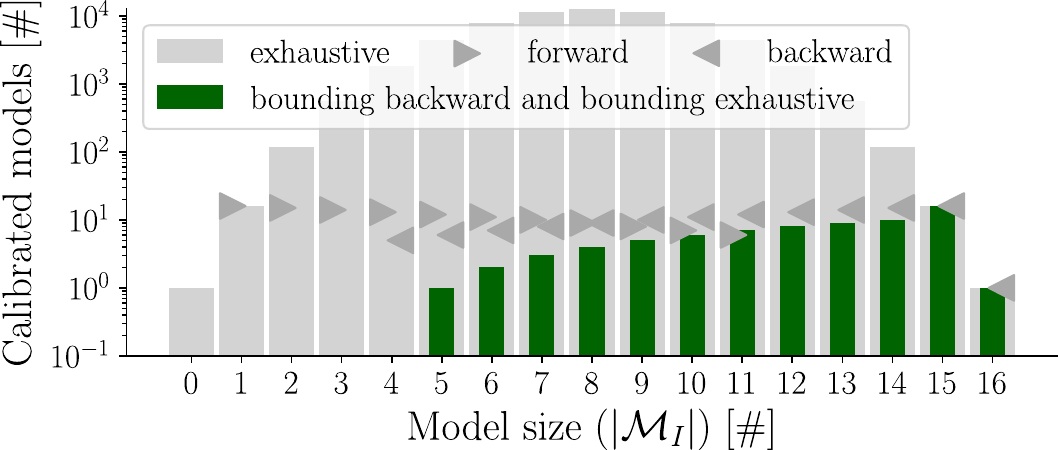}
\end{tabular}
\end{minipage}
\begin{minipage}[t]{.4\textwidth}
    \textbf{(f)}\\\hphantom{\textbf{(C)}}
    \includegraphics[scale=0.5, valign=t]{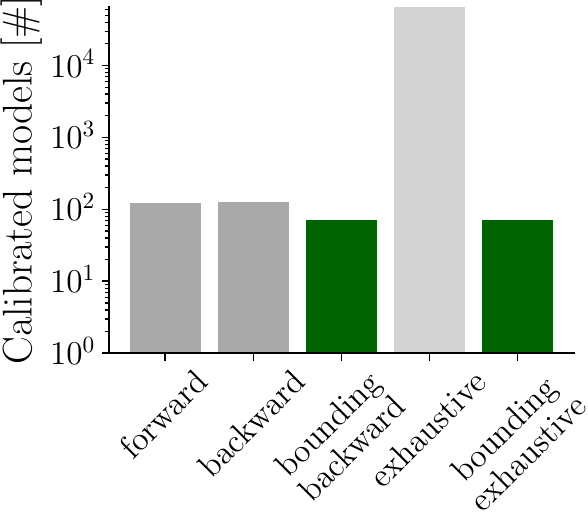}
\end{minipage}

    \caption{\textbf{Comparison of methods with a large-scale synthetic model selection problem}
    \textbf{(a)}~Schematic of the superset model, with all possible processes (arrows), and true model used for synthetic data.
    \textbf{(b)}~Synthetic data and maximum likelihood fit of the true model.
    \textbf{(c)}~Number of calibrated models at each model size, for each method.
    \textbf{(d)}~$\Delta\AIC$ values for the models returned by the each method. $\Delta\AIC$ is the difference to the optimal model $\AIC$, i.e. $\Delta\AIC = \AIC(\MoptCurr,\D) - \AIC(\Mopt,\D)$.
    \textbf{(e)}~Calibration time with each method.
    \textbf{(f)}~Number of calibrated models with each method.
    }
    \label{fig:famos_problem}
\end{figure}

To compare the performance of the standard and proposed methods in a controlled setup, we studied a previously published, synthetic model selection problem \citep{Gabel2019}. The model space is given by 4-compartment ODE models describing the abundance of species in the presence of self-replication and conversion (Figure~\ref{fig:famos_problem}a). As there are 16 different processes to choose from, the indicator vector is 16-dimensional, leading to a model space with $2^{16} = 65,536$ models. Following the original publication, slightly noisy synthetic data were generated with the true model (Figure~\ref{fig:famos_problem}b).

We performed model selection using the AIC and employed the forward, backward, and exhaustive brute-force methods, as well as the \bound{} backward and \bound{} exhaustive methods. The \bound{} exhaustive method employed the \bound{} backward selection for local searches. As a reference solution, we considered the result for the brute-force method, which calibrated all $65,536$ models. 

The optimal model found using the brute-force method matched the true model used for synthetic data generation. The AIC value of the true model was 231.58.  The true model was also recovered by standard and \bound{} backward selection, as well as \bound{} exhaustive search (Figure~\ref{fig:famos_problem}d). Only forward selection did not identify the optimal model.

The brute-force method required a computation time of $\approx 250.6$ CPU hours (Figure \ref{fig:famos_problem}e). All other search methods were substantially faster: the forward selection calibrated 121 models and used $\approx 0.3$ CPU hours; the backward selection calibrated 127 models and used $\approx 1.1$ CPU hours; and both the \bound{} backward and \bound{} exhaustive methods calibrated 72 models and used $0.6$ CPU hours. The higher computation time with the \bound{} variants, compared to the forward selection method, is a direct result of the (on average) higher complexity of the calibrated models, as noted earlier (Figures \ref{fig:famos_problem} C-F). Interestingly, the \bound{} exhaustive method behaved identically to the \bound{} backward selection because the first local search identified the optimal model and enabled exclusion of all (65,474) uncalibrated models, i.e., a 99.9\% reduction in the number of calibrated models compared to the brute-force method.

The assessment of the complexity of the calibrated models revealed that, with the \bound{} methods, most models with fewer than 15 parameters were excluded by the condition \eqref{eq:condition} (Figure~\ref{fig:famos_problem}c). While $100\%$ of the models with 15 and 16 parameters were calibrated, only $8.3\%$ of the models with 14 parameters and $<0.055\%$ of all models with fewer than 12 parameters were calibrated. Furthermore, although the models with up to 14 parameters ($|\Mcand| \leq 14$) constitute 99.97\% of the model space, only $0.084\%$ of them needed to be calibrated to achieve the same result as the brute-force method.

In summary, the proposed search methods achieved a better performance than the state-of-the-art. In particular, to identify the globally best model, the computation time was reduced by $99.7$\% and the number of calibrated models by $99.9$\%.

\subsection*{\BOUND{} methods outperform standard methods in applications}

To determine if the superior performance observed for the synthetic problems hold for real applications, we studied two previously-published real-world problems from different domains.

\textbf{(1) The Hald dataset on the heat of hardening of cement \cite{Hald1952}}

Hald \cite{Hald1952} described a cement dataset and a corresponding linear regression variable selection problem which has been widely studied in the statistical literature. The heat of hardening of cement ($y \in \mathbb{R}$) is modelled as a linear function of the abundances of four main cement ingredients ($x \in \mathbb{R}_+^4$):
\begin{equation*}
y = \theta^T x + \theta_0 + \varepsilon, \qquad \varepsilon \sim \mathcal{N}(0,\sigma^2),
\end{equation*}
with measurement noise $\varepsilon \in \mathbb{R}$. The model space is generated by taking variations of the ingredients in the model, i.e., the elements of $\theta \in \mathbb{R}^4$ can individually be set to zero, yielding $2^{4}=16$ candidate models. We applied the forward, backward, \bound{} backward, brute-force, and \bound{} exhaustive methods (with \bound{} backward local searches).

\begin{figure}[!ht]
\hspace{0.5cm}
\begin{minipage}[t]{.4\textwidth}
\textbf{(a)}\\\hphantom{()..}
    \includegraphics[scale=0.5, valign=t]{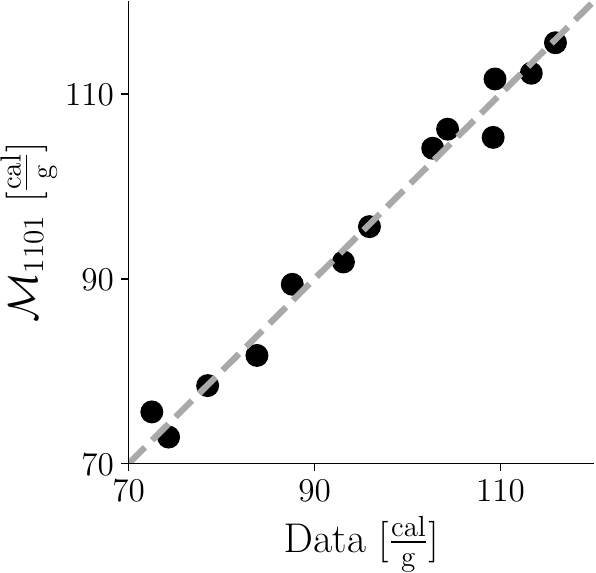}
\end{minipage}
\begin{minipage}[t]{.6\textwidth}
    \textbf{(b)}\\\hphantom{\textbf{(C)}}
    \includegraphics[scale=0.5, valign=t]{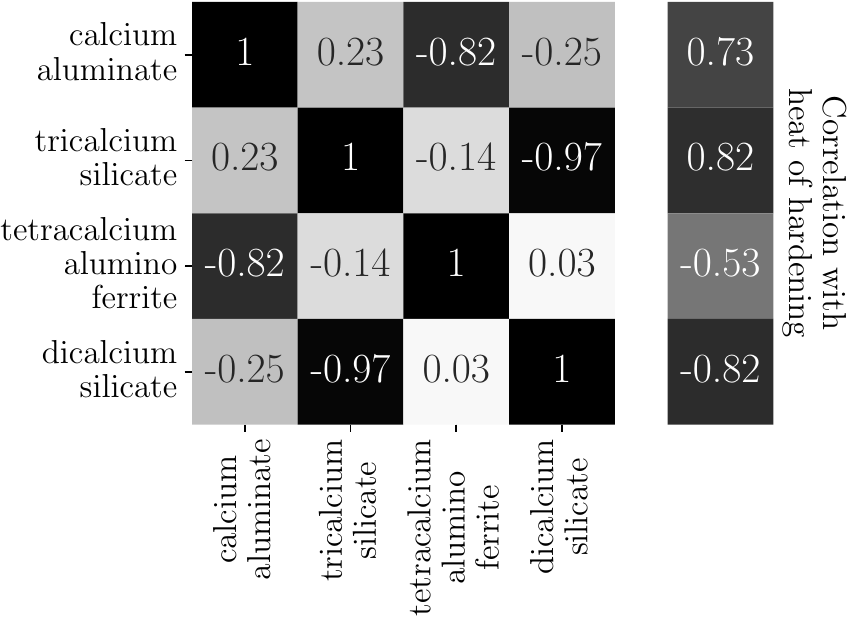}
\end{minipage}

\hspace{0.5cm}
\begin{minipage}[t]{.4\textwidth}
\textbf{(c)}\\\hphantom{\textbf{(C)}}
    \includegraphics[scale=0.5, valign=t]{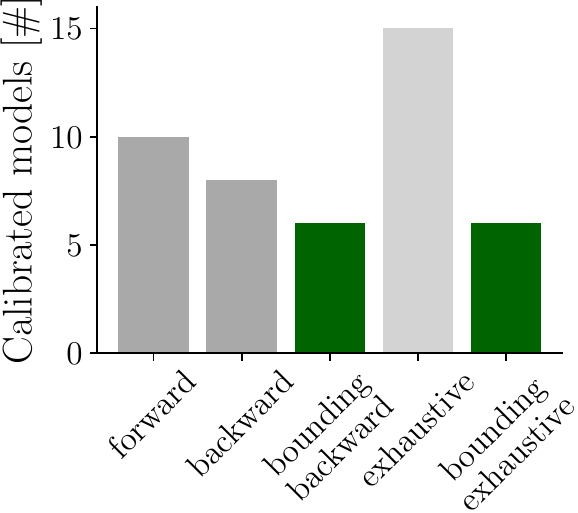}
\end{minipage}
\begin{minipage}[t]{.7\textwidth}
    \textbf{(d)}\\\hphantom{\textbf{(C)}}
    \includegraphics[width=0.6\textwidth, valign=t]{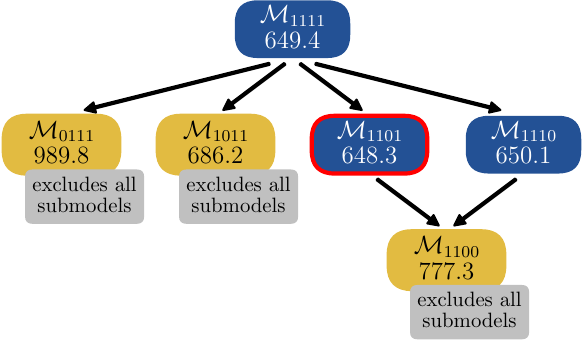}
\end{minipage}

    \caption{\textbf{Application example with linear regression and Hald's cement dataset}
    \textbf{(a)}~Optimal model ($\M_{1101}$) fit vs. data of the heat of hardening.
    \textbf{(b)}~Pearson correlation of the cement ingredients abundances, in the dataset. Superscript numbers indicate the position of ingredients effect parameters in the model indicator vector $I$.
    \textbf{(c)}~Histogram of the number of calibrated models with each method.
   \textbf{(d)}~Model space search path with the \bound{} backward selection.}
    \label{fig:hald_problem}
\end{figure}

With the AIC as the model selection criterion, all methods identified $\M_{1101}$ as the best model, and $\M_{1101}$ provides a good description of the experimental data (Figure~\ref{fig:hald_problem}a). The results with the $\AICc$, $\BIC$, and Mallows's $\Cp$ are provided in the \supp{}. The absence of the third ingredient abundance from the model is plausible because: it is highly correlated with the first ingredient; it is least correlated with the heat of hardening (Figure~\ref{fig:hald_problem}b); and there is some collinearity because these four main ingredients of cement sum up to nearly 100\% of the cement composition (see \supp{}).

Each of the methods calibrated between $37.5\%$ and $100\%$ of all 16 models (Figure~\ref{fig:hald_problem}c). The \bound{} backward and \bound{} exhaustive methods calibrated the fewest models (6) and shared an identical search path (Figure~\ref{fig:hald_problem}d).

\textbf{(2) The Blasi et al. dataset on the acetylation states of histone H4 \cite{Blasi2016}}

\begin{figure}[!ht]
\begin{tabular}[t]{l}
    \textbf{(a)}\enskip%\\% AIC values.\\
    \includegraphics[width=0.8\textwidth, valign=t]{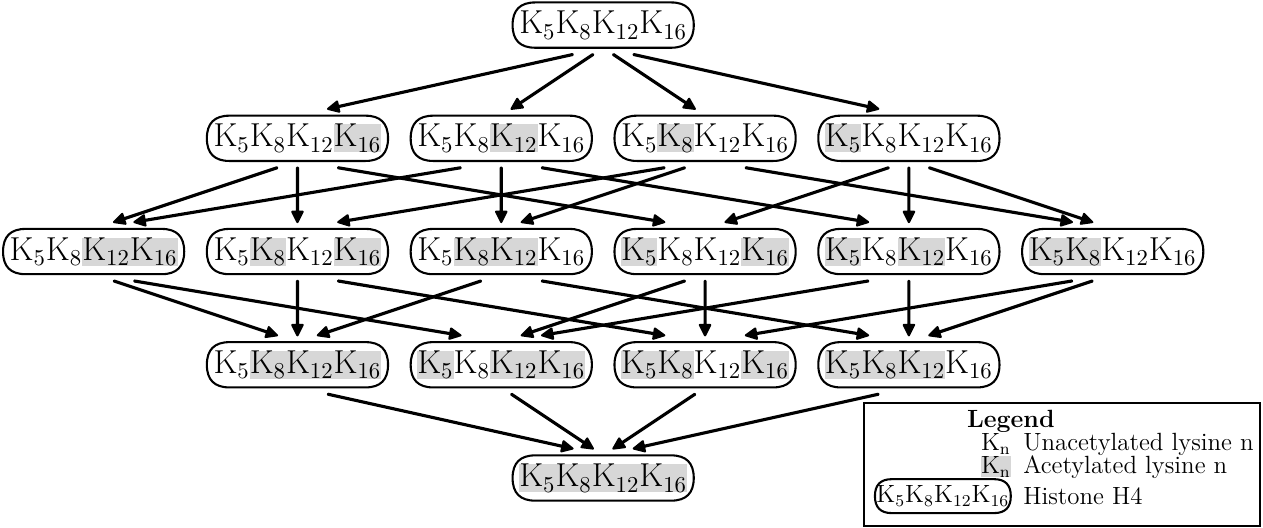}
\end{tabular}
\vspace{0.5cm}\\
\begin{tabular}[t]{@{}l@{}l}
\begin{tabular}[t]{l}
    \textbf{(b)}\enskip\\%\hphantom{0.1cm}% Size of all models.\\
    \includegraphics[scale=1, valign=t]{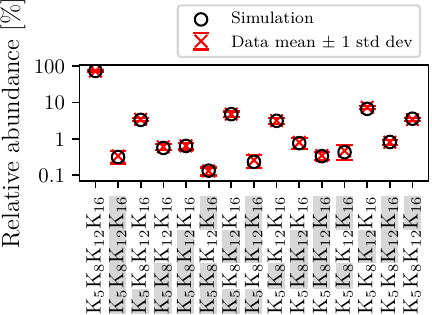}
\end{tabular}
&\begin{tabular}[t]{l}
    \textbf{(c)}\enskip\\%\hphantom{0.1cm}%\\% \# calibrated models.\\
    \includegraphics[scale=0.6, valign=t]{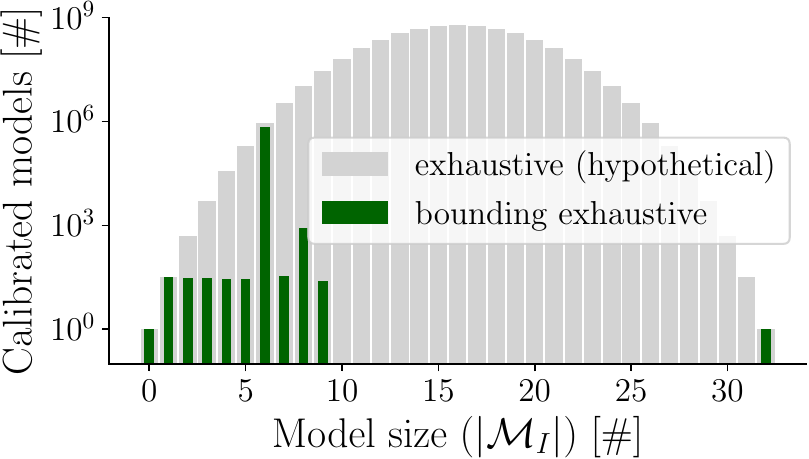}
\end{tabular}
\end{tabular}

    \caption{\textbf{Application example with ODE modeling of Blasi et al.'s histone acetylation dataset}
    \textbf{(a)}~Illustration of the topology of the mathematical model for histone acetylation. The model selection problem is to identify which of these processes (arrows) occurs with a rate constant different to the default.
    \textbf{(b)}~Optimal model fit vs. data of the relative abundance of histone acetylation states.
    \textbf{(c)}~Number of calibrated models at each model size, for the \bound{} exhaustive method. The theoretical number for the brute-force exhaustive method is also displayed.}
    \label{fig:blasi_problem}
\end{figure}

Blasi et al. \cite{Blasi2016} described a prototypical nonlinear problem for model-based data analysis in biology and biochemistry. They introduced an ODE model for the abundance of 16 histone acetylation motifs (Figure~\ref{fig:blasi_problem}a), where all acetylation rates proceeded with the same rate constant by default. The model selection problem was to determine which of the rate constants are motif-specific; i.e., which rate constants are different to the default rate constant. Overall, there are 32 independent design variables (possibly non-default rate constants), yielding a model space with $2^{32}\approx 4$ billion candidate models. To ensure the reliability of the results and account for the data scarcity, we used the $\AICc$ to identify the globally-optimal model.

As the ODE system in this problem is linear, the original publication derived the closed-form analytical solution and used that efficient representation of the model for a brute-force search. However, to ensure this example is representative of model selection problems involving nonlinear, dynamical descriptions of biology more generally, we did not exploit this model property, which meant the brute-force method was infeasible. Accordingly, a reduction in the search space is particularly beneficial. As the degree of search space reduction achieved by the proposed \bound{} methods depends on the best criterion seen so far $\AICc(\MoptCurr)$, we employed a \bound{} exhaustive approach where the first local search was the FAMoS method, and subsequent local searches were the \bound{} backward selection, as motivated earlier.

For the considered problem, the initial FAMoS search calibrated 1,077 models, and yielded a model $\MoptCurr$ with 7 non-default rate constants, which provided a good fit of the data (Figure~\ref{fig:blasi_problem}b). The bound and condition \eqref{eq:condition} excluded 99.98\% (4,294,095,768) of the model space, leaving 870,451 uncalibrated models that could potentially improve on $\MoptCurr$. The remaining models were calibrated or excluded according to \bound{} backward searches. The \bound{} exhaustive method terminated after 710,655 model calibrations, and identified the best model provided by FAMoS as the globally optimal model. This optimal model matched the optimal model reported in the original publication. Interestingly, the use of FAMoS as the initial local search method resulted in a calibrated model size distribution (Figure~\ref{fig:blasi_problem}c) that differed from that produced when the \bound{} backward selection was the initial local search method in the large-scale problem (Figure \ref{fig:famos_problem}c). All uncalibrated models of size $|\M_I|\geq7$ were excluded after the initial FAMoS search using the bound on the full models, while many candidates with $|\M_I| \leq 6$ required calibration.

The \bound{} exhaustive method enabled a substantial improvement in computation time as well. While the \bound{} exhaustive method took 26.9 CPU days to complete, we conservatively estimate that the brute-force method would take $\approx \frac{26.9}{365} \frac{2^{32}}{710,655} \approx 445$ CPU years. This estimate is conservative because it is based on the calibration time of models encountered by the \bound{} exhaustive method, which mostly possessed a small number of parameters ($|\M_I|\leq 6$).

\section*{Discussion}

In this contribution, we describe (i) simple bounds on the criterion values, for the non-monotonic criteria AIC, $\AICc$, BIC, and Mallows's $\Cp$, of uncalibrated models, and (ii) conditions that can be evaluated during model or feature selection \textit{a priori} model calibration to exclude models. We (iii) develop \bound{} variants of popular methods, such as backward and brute-force searches, that are similar to pre-existing branch-and-bound methods for monotonic criteria, and (iv) demonstrate these \bound{} methods on various kinds of synthetic and real problems. Theoretically, the \bound{} methods can exclude more models if a good model is found quickly, and if models that have many submodels are calibrated.

We evaluated the performance of the proposed \bound{} methods on two synthetic examples and two real-world application problems. In all cases, the \bound{} methods identified the same optimal model at a lower computational cost. In the large real-data problem from Blasi et al. \citep{Blasi2016}, 99.98\% of the model space was excluded without calibration, significantly reducing the computational cost of an exhaustive assessment of all models. Thus, the \bound{} methods can render model selection problems, for which only local methods were feasible, amenable to a global search.

These proposed methods provide the same results as their established counterparts if model calibration guarantees the maximum likelihood estimate. However, many calibration problems are non-convex, making such guarantees challenging. We successfully used multi-start methods for non-convex calibration problems, but this still does not provide a guarantee. It would be interesting to investigate how the proposed and pre-existing methods perform when individual calibrations yield suboptimal results. Standard and proposed local search methods might stop prematurely, and standard and proposed exhaustive search might yield a suboptimal model. In theory, this would lead to exclusion by the \bound{} methods of potentially better submodels, according to the condition \eqref{eq:condition}. Although no model selection method can be robust to suboptimal model calibration, the \bound{} exhaustive (and standard brute-force) can no longer guarantee that the optimal model is returned; only the optimal model up to model calibration. Overall, a study of the impact of suboptimal calibration on model selection would be valuable in general.

Beyond this, we see the following future research directions.
(i) Incorrect rejection of a model due to suboptimal calibration should be avoided. While the allocation to each model of the same number of multi-start optimiser runs proved sufficient in this study, a more dynamic allocation of computational resources depending on calibration problem characteristics might be beneficial. For example, larger models (larger $|\M_I|$) tend to be more computationally costly to calibrate.
(ii) The better the best model seen so far $\MoptCurr$, the more models that can be excluded according to the condition for the optimal model \eqref{eq:condition}. In this study, we used the efficient metaheuristics provided by FAMoS to identify good models quickly. However, it would be beneficial to explore additional local search methods, in particular methods based on L1-regularization \citep{SteiertTim2016}.
(iii) For large model spaces, storing candidate models can be challenging. Our current implementation for the model by Blasi et al. \citep{Blasi2016} requires $2^{32}$ bits, or $\approx$0.5 GB, of memory. A more efficient encoding, and more tailored implementations of the \bound{} methods, would be beneficial. For example, if all submodels of some model can be excluded, only this information need be stored in memory, not the explicit ``excluded?'' bit for each submodel. 
(iv) Calibration of suboptimal models may provide additional exclusions. In the Blasi et al. \citep{Blasi2016} example, after the initial FAMoS local search and corresponding exclusions, 1,077 model calibrations removed all with 7 or more parameters. The subsequent \bound{} backward searches involved calibration of 709,577 models with 6 parameters. This could be reduced by identifying and calibrating a relative superset of many of these models with 6 parameters, and thereby exclude many of these 709,577 models according to the condition \eqref{eq:condition}. The parameters to include in this relative superset could be informed by a cheap identifiability analysis, such as the Fisher Information Matrix.

In this study, we focused on identifying the best model. In many applications, it is important to find not just the best model, but all models that cannot be rejected according to some threshold \citep{Burnham2004}. Adapting our algorithm to compute and evaluate the set of plausible models is straightforward, by replacing the condition for the optimal model \eqref{eq:condition} with the condition for all plausible models ((1) in \supp{}). These plausible models can then be used for model averaging, using AIC- or BIC-weights. An interesting extension would be to collect the plausible models such that a certain fraction of the weight (which sums to 1 over the model space) is contained in the set of plausible models. This introduces a dependency on the current set of plausible models, in the condition.

In summary, despite their simplicity, the bounds on information criteria hold great potential for model selection across application areas. Model selection for problems with many design variables could be greatly simplified. The availability of our implementation in the Python package \texttt{modelSelector} under a permissive license model ensures ease-of-use and facilitates future extensions.

\section*{Methods}

\subsection*{Model selection methods}
The \bound{} methods typically require access to the set of all models $\Mall$, the set of all calibrated models $\Mcal$, and the set of all excluded models $\Mex$, the set of all unseen models $\Munseen := \Mall \setminus (\Mcal \cup \Mex)$. Calibration implies that the criterion value was computed. All criterion values are implicitly stored in memory once computed, to avoid recomputation, e.g. when checking the condition \eqref{eq:condition}.

The set of all excludable models $\Mexable$ is the subset of all unseen models $\Munseen$ that can be excluded by any means, e.g. with either the bound on the criterion $\Bound_\Crit(\Msub, \Mcand)$ or the bound on the model size $\nBound_\Crit(\Msub, \Mcand)$, via either the condition for the optimal model \eqref{eq:condition} or the condition for all plausible models ((1) in \supp{}).

The algorithms here are simplified for presentation and can be considered canonical but not optimised -- there are several improvements that can be made to reduce the computational complexity, some of which are implemented in the \texttt{ModelSelector} and \texttt{PEtab Select} packages.

A generic algorithm for a \bound{} local method is provided by Algorithm \eqref{alg:bounded_local}, which enables the bounding feature for stepwise methods such as forward and backward selection. It involves the termination criterion $\Aterm$, and the set of plausible models for the current iteration $\Aplaus$.

\begin{algorithm}[ht]
\caption{\BOUND{} local method}\label{alg:bounded_local}
\begin{algorithmic}[1]

\Function{\BOUND{}Local}{}
    \State $\Mnext \gets$ method-appropriate (or user-supplied) calibrated initial model $\M_0$.
    \State $\Mcal \gets \{\Mnext \}$, $\Mex \gets \emptyset$, $\MoptCurr \gets $ NaN.

    \While{$\neg \Aterm$}
        \State $\MoptCurr \gets \Mnext$. Set $\Aplaus$ according to the method.
        \For {$\Mcand \in \left\{ \M : \M \in \Munseen \wedge \M \in \Aplaus \right\}$}
            \State Calibrate $\Mcand$, add $\Mcand$ to $\Mcal$.
            \State Compute $\Mexable$, add $\Mexable$ to $\Mex$. \label{alg:bounded_local:bounding}
            \State $\Mnext \gets \argmin_\M \left\{\Crit(\M) : \M \in \Mcal\right\}$
        \EndFor
    \EndWhile
    \State \Return $\MoptCurr$ 
\EndFunction
\end{algorithmic}
\end{algorithm}

For example, the \bound{} backward selection is defined by starting at the superset model $\M_0 := \Msup$, considering models that have $r$ fewer parameters than the best model seen so far $\Aplaus := \left\{\M : |\M| = |\MoptCurr| - r\right\}$, and terminating when an iteration does not improve on the previous iteration $\Aterm := \MoptCurr = \Mnext$. Typically, $r=1$ but, if $\Aplaus = \emptyset$ with $r=1$, then $r$ can be successively incremented until either $\Aplaus \neq \emptyset$ or $r = |\MoptCurr|$.

The algorithms for standard (non-\bound{}) local methods are equivalent to the \bound{} local methods, except $\Mexable$ (and $\Mex$) are always empty. For example, the standard backward selection is equivalent to the \bound{} backward selection, except Line \ref{alg:bounded_local:bounding} is skipped.

The FAMoS method uses metaheuristics that substantially complicate the local search. Interested readers can refer to the original publication \citep{Gabel2019} for a full description and graphical representation. The original publication also provides a canonical implementation in R, available at \url{https://cran.r-project.org/package=FAMoS}. A Python implementation is available in \texttt{PEtab Select} \citep{Pathirana2024}.

The \bound{} exhaustive method can be implemented with many variations. The approach taken here is to use a sequence of local searches generated by some function $\mathcal{S}:\mathbb{N} \rightarrow \text{LocalMethod}$ until all models are either calibrated or excluded. LocalMethod represents, for example, the \bound{} backward selection. This approach is presented in Algorithm \ref{alg:bounded_global}.

\begin{algorithm}[ht]
\caption{\BOUND{} exhaustive method}\label{alg:bounded_global}
\begin{algorithmic}[1]

\Function{\BOUND{}BruteForce}{}
    \State $\text{MethodIndex} \gets 0$.
    \State $\Mcal \gets \emptyset$, $\Mex \gets \emptyset$, $\MoptCurr \gets $ NaN.
    \State $\Mcal, \Mex, \MoptCurr$ are shared with all local methods.
    \While{$\Munseen \neq \emptyset$}
        \State Run $\mathcal{S}(\text{MethodIndex})$, update $\Mcal$, $\Mex$, and $\MoptCurr$.
        \State Increment MethodIndex
    \EndWhile
    \State \Return $\MoptCurr$ 
\EndFunction
\end{algorithmic}
\end{algorithm}

\subsection*{Implementation}
Model selection methods were implemented in the \texttt{modelSelector} and \texttt{PEtab Select} Python packages. The criteria AIC, corrected AIC, and BIC can be used with both packages. \texttt{modelSelector} additionally supports Mallows's $\Cp$ and the \bound{} methods, except the \bound{} FAMoS method. Both packages support the backward, forward, and brute-force methods. The \texttt{PEtab Select} package additionally supports the FAMoS method.

The \texttt{modelSelector} package provides direct support for linear and logistic regression models via \texttt{scikit-learn} \citep{scikit-learn}, as well as an interface for using custom models and objective functions. For ODE models, a specification using \texttt{PEtab} \citep{SchmiesterSch2021} is supported, including a formulation of models in the SBML format \citep{Hucka2019} via \texttt{yaml2sbml} \citep{Vanhoefer2021}. In this case, \texttt{AMICI} \citep{Frohlich2021} is used for model simulation and sensitivity analysis, while \texttt{pyPESTO} is employed for parameter estimation \citep{Schaelte2023}. The optimization is performed using multi-start local optimization with the \texttt{Fides} optimiser \citep{Frohlich2022}, with 20 multi-starts initialised at uniformly-random starting points.

The \texttt{PEtab Select} package was used to perform model selection with the FAMoS, backward, and brute-force methods, and with \texttt{AMICI}, \texttt{Fides}, \texttt{PEtab}, and \texttt{pyPESTO} as above.

\section*{Data availability}
All data used to perform the model selection and produce the visualizations are available on Zenodo (\url{https://doi.org/10.5281/zenodo.14906119}).

\section*{Code availability}
All packages mentioned in the \textit{Methods} subsection ``Implementation'' are open-source. The \texttt{modelSelector} package can be installed from GitHub (\url{https://github.com/ICB-DCM/modelSelector}), and all other packages can be installed from the Python Package Index (\url{https://pypi.org}). All additional code used to perform the model selection and produce the visualizations are available on Zenodo (\url{https://doi.org/10.5281/zenodo.14906119}).

\section*{Acknowledgements}
We are grateful to Vincent Wieland, for providing an implementation of the model selection problem, and corresponding parameter estimation problems, for the illustrative small-scale model selection problem. Computations were performed on the Bonna and Unicorn clusters at the University of Bonn.

This work was supported by the TRA Life and Health (University of Bonn) as part of the Excellence Strategy of the federal and state governments, the Deutsche Forschungsgemeinschaft (DFG, German Research Foundation) under Germany’s Excellence Strategy (project IDs 390685813 - EXC 2047 and 390873048 - EXC 2151) and through Metaflammation, project ID 432325352 – SFB 1454, and by the University of Bonn via the Schlegel professorship to J.H.

\section*{Author contributions}
JV, DP and JH designed the study, JV, AK and DP implemented the examples and analysed the data and proved the inequalities. JV, AK, DD and DP implemented the algorithms. JV wrote the \texttt{modelSelector} package. All authors wrote the manuscript and approved the final version.

\textbf{Contributor Role Taxonomy (CRediT)}.
Conceptualization: DP, JH, JV.
Data curation: AK, DP, JV.
Formal analysis: JV, AK, DP.
Funding acquisition: DP, JH, JV.
Investigation: AK, DP, JV.
Methodology: AK, DD, DP, JV.
Project administration: DP, JH, JV.
Software: AK, DD, DP, JV.
Visualization: DP, JH, JV.
Writing – original draft: DP, JH, JV.
Writing – review \& editing: AK, DD, DP, JH, JV.

\section*{Declaration of interests}
The authors declare no competing interests.

\section*{Resource availability}
\paragraph{Lead contact}
Requests for further information and resources should be directed to and will be fulfilled by the lead contact, Dilan Pathirana (dilan.pathirana@uni-bonn.de).

\paragraph{Materials availability}
This study did not generate new unique reagents.

\paragraph{Data and code availability}
\begin{itemize}
    \item All data have been deposited at Zenodo and are publicly available as of the date of publication at \url{https://doi.org/10.5281/zenodo.14906119}.
    \item All original code has been deposited at Zenodo and is publicly available at \url{https://doi.org/10.5281/zenodo.14906119} as of the date of publication.
    \item Any additional information required to reanalyze the data reported in this paper is available from the lead contact upon request.
\end{itemize}

\clearpage
\newpage
\printbibliography

@book{Jeffreys1961,
	address = {Oxford},
	author = {Jeffreys, H.},
	edition = {3rd},
	publisher = {Oxford University Press},
	title = {Theory of Probability},
	year = {1961}}

@inproceedings{Akaike1973,
    title={Information theory and an extension of the maximum likelihood principle},
    author={Akaike, H.},
    booktitle={2nd International Symposium on Information Theory},
    year={1973},
    pages={267--281}, 
    doi={doi:10.1007/978-1-4612-1694-0_15}
}

@article{Burnham2004,
  title={Model selection and multimodel inference},
  author={Burnham, Kenneth P and Anderson, David R},
  journal={A practical information-theoretic approach},
  volume={2},
  year={2004},
  doi={doi:10.1007/b97636}
}

@article{Hurvich1989,
  title={Regression and time series model selection in small samples},
  author={Hurvich, Clifford M and Tsai, Chih-Ling},
  journal={Biometrika},
  volume={76},
  number={2},
  pages={297--307},
  year={1989},
  publisher={Oxford University Press}, 
  doi = {doi:10.1093/biomet/76.2.297}
}

@article{Schwarz1978,
  title={Estimating the dimension of a model},
  author={Schwarz, Gideon},
  journal={The Annals of Statistics},
  pages={461--464},
  year={1978}, 
  doi = {doi:10.1214/aos/1176344136}
}

@article{Mallows1973,
  title={Some comments on Cp},
  author={Mallows, Colin L},
  journal={Technometrics},
  volume={15},
  number={4},
  pages={661--675},
  year={1973},
  publisher={Taylor \& Francis}, 
  doi = {doi:10.1080/00401706.1973.10489103}
}

@book{Hald1952,
  title={Statistical Theory with Engineering Applications},
  author={Hald, A.},
  series={A Wiley publication in applied statistics},
  publisher={John Wiley \& Sons},
  year={1952}
}

@article{Gabel2019,
  title={FAMoS: A Flexible and dynamic Algorithm for Model Selection to analyse complex systems dynamics},
  author={Gabel, Michael and Hohl, Tobias and Imle, Andrea and Fackler, Oliver T and Graw, Frederik},
  journal={PLOS Computational Biology},
  volume={15},
  number={8},
  pages={e1007230},
  year={2019},
  publisher={Public Library of Science San Francisco, CA USA}, 
  doi = {doi:10.1371/journal.pcbi.1007230}
}

@article{Blasi2016,
  title={Combinatorial histone acetylation patterns are generated by motif-specific reactions},
  author={Blasi, Thomas and Feller, Christian and Feigelman, Justin and Hasenauer, Jan and Imhof, Axel and Theis, Fabian J and Becker, Peter B and Marr, Carsten},
  journal={Cell systems},
  volume={2},
  number={1},
  pages={49--58},
  year={2016},
  publisher={Elsevier}, 
  doi = {doi:10.1016/j.cels.2016.01.002}
}

@article{Hucka2019,
    title = {The Systems Biology Markup Language (SBML): Language Specification for Level 3 Version 2 Core Release 2},
    author = {Michael Hucka and Frank T. Bergmann and Claudine Chaouiya and Andreas Dräger and Stefan Hoops and Sarah M. Keating and Matthias König and Nicolas Le Novère and Chris J. Myers and Brett G. Olivier and Sven Sahle and James C. Schaff and Rahuman Sheriff and Lucian P. Smith and Dagmar Waltemath and Darren J. Wilkinson and Fengkai Zhang},
    pages = {20190021},
    volume = {16},
    number = {2},
    journal = {Journal of Integrative Bioinformatics},
    doi = {doi:10.1515/jib-2019-0021},
    year = {2019}
}

@article{Vanhoefer2021,
  year = {2021},
  publisher = {The Open Journal},
  volume = {6},
  number = {61},
  pages = {3215},
  author = {Jakob Vanhoefer and Marta R. A. Matos and Dilan Pathirana and Yannik Schälte and Jan Hasenauer},
  title = {yaml2sbml: Human-readable and -writable specification of ODE models and their conversion to SBML},
  journal = {Journal of Open Source Software}
}

@article{SteiertTim2016,
	author = {Steiert, Bernhard and Timmer, Jens and Kreutz, Clemens},
	doi = {10.1093/bioinformatics/btw461},
	journal = {Bioinformatics},
	number = {17},
	pages = {i718--i726},
	publisher = {Oxford Univ Press},
	title = {L1 regularization facilitates detection of cell type-specific parameters in dynamical systems},
	volume = {32},
	year = {2016}}

@article{SchmiesterSch2021,
  author    = {Schmiester, Leonard AND Schälte, Yannik AND Bergmann, Frank T. AND Camba, Tacio AND Dudkin, Erika AND Egert, Janine AND Fröhlich, Fabian AND Fuhrmann, Lara AND Hauber, Adrian L. AND Kemmer, Svenja AND Lakrisenko, Polina AND Loos, Carolin AND Merkt, Simon AND Müller, Wolfgang AND Pathirana, Dilan AND Raimúndez, Elba AND Refisch, Lukas AND Rosenblatt, Marcus AND Stapor, Paul L. AND Städter, Philipp AND Wang, Dantong AND Wieland, Franz-Georg AND Banga, Julio R. AND Timmer, Jens AND Villaverde, Alejandro F. AND Sahle, Sven AND Kreutz, Clemens AND Hasenauer, Jan AND Weindl, Daniel},
  journal   = {PLOS Computational Biology},
  title     = {PEtab—Interoperable specification of parameter estimation problems in systems biology},
  year      = {2021},
  number    = {1},
  pages     = {1-10},
  volume    = {17},
  publisher = {Public Library of Science},
}

@software{Pathirana2024,
  author       = {Dilan Pathirana and
                  Daniel Weindl and
                  Yannik Schälte and
                  Domagoj Doresic},
  title        = {PEtab-dev/petab\_select: PEtab Select v0.2.1},
  year         = 2024,
  publisher    = {Zenodo},
  version      = {v0.2.1},
  doi          = {10.5281/zenodo.14184839},
  url          = {https://zenodo.org/records/14184839},
}

@article{Schaelte2023,
      author = {Schälte, Yannik and Fröhlich, Fabian and Jost, Paul J and Vanhoefer, Jakob and Pathirana, Dilan and Stapor, Paul and Lakrisenko, Polina and Wang, Dantong and Raimúndez, Elba and Merkt, Simon and Schmiester, Leonard and Städter, Philipp and Grein, Stephan and Dudkin, Erika and Doresic, Domagoj and Weindl, Daniel and Hasenauer, Jan},
    title = "{pyPESTO: a modular and scalable tool for parameter estimation for dynamic models}",
    journal = {Bioinformatics},
    volume = {39},
    number = {11},
    year = {2023},
}

@article{Frohlich2022,
  title={Fides: Reliable trust-region optimization for parameter estimation of ordinary differential equation models},
  author={Fr{\"o}hlich, Fabian and Sorger, Peter K},
  journal={PLoS Computational Biology},
  volume={18},
  number={7},
  pages={e1010322},
  year={2022},
  publisher={Public Library of Science San Francisco, CA USA}, 
  doi={doi:10.1371/journal.pcbi.1010322}
}

@article{scikit-learn,
 title={Scikit-learn: Machine Learning in {P}ython},
 author={Pedregosa, F. and Varoquaux, G. and Gramfort, A. and Michel, V. and Thirion, B. and Grisel, O. and Blondel, M. and Prettenhofer, P. and Weiss, R. and Dubourg, V. and Vanderplas, J. and Passos, A. and Cournapeau, D. and Brucher, M. and Perrot, M. and Duchesnay, E.},
 journal={Journal of Machine Learning Research},
 volume={12},
 pages={2825--2830},
 year={2011},
 doi={doi:10.48550/arXiv.1201.0490}
}

@article{Frohlich2021,
  title={AMICI: High-Performance Sensitivity Analysis for Large Ordinary Differential Equation Models},
  author={Fr{\"o}hlich, Fabian and Weindl, Daniel and Sch{\"a}lte, Yannik and Pathirana, Dilan and Paszkowski, {\L}ukasz and Lines, Glenn Terje and Stapor, Paul and Hasenauer, Jan},
  journal = {Bioinformatics},
  year = {2021},
}

@article{Calcagno2010,
  title={glmulti: an R package for easy automated model selection with (generalized) linear models},
  author={Calcagno, Vincent and de Mazancourt, Claire},
  journal={Journal of statistical software},
  volume={34},
  pages={1--29},
  year={2010},
  doi={10.18637/jss.v034.i12}
}

@book{Grafen2002,
  title={Modern statistics for the life sciences},
  author={Grafen, Alan and Hails, Rosemary},
  year={2002},
  publisher={Oxford University Press}
}

@article{Kirk2013,
  title={Model selection in systems and synthetic biology},
  author={Kirk, Paul and Thorne, Thomas and Stumpf, Michael PH},
  journal={Current opinion in biotechnology},
  volume={24},
  number={4},
  pages={767--774},
  year={2013},
  publisher={Elsevier}
}

@article{Ding2018,
  title={Model selection techniques: An overview},
  author={Ding, Jie and Tarokh, Vahid and Yang, Yuhong},
  journal={IEEE Signal Processing Magazine},
  volume={35},
  number={6},
  pages={16--34},
  year={2018},
  publisher={IEEE}
}

@article{Kitano2002,
  title={Systems biology: a brief overview},
  author={Kitano, Hiroaki},
  journal={Science},
  volume={295},
  number={5560},
  pages={1662--1664},
  year={2002},
  publisher={American Association for the Advancement of Science}
}

@book{Mesterton2011,
  title={A concrete approach to mathematical modelling},
  author={Mesterton-Gibbons, Mike},
  year={2011},
  publisher={John Wiley \& Sons}
}

@misc{BenchmarkingInitiativeBenchmarkModelsPEtabBenchmark,
  author = {{The PEtab Benchmark Collection contributors}},
  title = {{The PEtab Benchmark Collection of parameter estimation problems}},
  year = 2024,
  publisher = {Zenodo},
  version = {2024.10.15},
  doi = {10.5281/zenodo.8155057},
  url = {https://zenodo.org/records/13963472},
}

@article{Saltelli2019,
  title={A short comment on statistical versus mathematical modelling},
  author={Saltelli, Andrea},
  journal={Nature communications},
  volume={10},
  number={1},
  pages={3870},
  year={2019},
  publisher={Nature Publishing Group UK London}
}

@article{Bodner2021,
  title={Ten simple rules for tackling your first mathematical models: A guide for graduate students by graduate students},
  author={Bodner, Korryn and Brimacombe, Chris and Chenery, Emily S and Greiner, Ariel and McLeod, Anne M and Penk, Stephanie R and Vargas Soto, Juan S},
  journal={PLoS Computational Biology},
  volume={17},
  number={1},
  pages={e1008539},
  year={2021},
  publisher={Public Library of Science San Francisco, CA USA}
}

@article{Villaverde2022,
  title={A protocol for dynamic model calibration},
  author={Villaverde, Alejandro F and Pathirana, Dilan and Fr{\"o}hlich, Fabian and Hasenauer, Jan and Banga, Julio R},
  journal={Briefings in Bioinformatics},
  volume={23},
  number={1},
  year={2022},
  publisher={Oxford University Press}
}

@article{Rodriguez2013,
  title={Simultaneous model discrimination and parameter estimation in dynamic models of cellular systems},
  author={Rodriguez-Fernandez, Maria and Rehberg, Markus and Kremling, Andreas and Banga, Julio R},
  journal={BMC systems biology},
  volume={7},
  number={1},
  pages={1--14},
  year={2013},
  publisher={BioMed Central}
}

@article{furnivalRegressionsLeapsBounds1974,
  title = {Regressions by {{Leaps}} and {{Bounds}}},
  author = {Furnival, George M. and Wilson, Robert W.},
  date = {1974-11-01},
  year={1974},
  journaltitle = {Technometrics},
  volume = {16},
  number = {4},
  pages = {499--511},
  publisher = {Taylor \& Francis},
  issn = {0040-1706},
  doi = {10.1080/00401706.1974.10489231}
}

@article{kohaviWrappersFeatureSubset1997,
  title = {Wrappers for Feature Subset Selection},
  author = {Kohavi, Ron and John, George H.},
  date = {1997-12-01},
  year={1997},
  journaltitle = {Artificial Intelligence},
  shortjournal = {Artificial Intelligence},
  series = {Relevance},
  volume = {97},
  number = {1},
  pages = {273--324},
  issn = {0004-3702},
  doi = {10.1016/S0004-3702(97)00043-X}
}

@article{narendraBranchBoundAlgorithm1977,
  title = {A {{Branch}} and {{Bound Algorithm}} for {{Feature Subset Selection}}},
  author = {{Narendra} and {Fukunaga}},
  date = {1977-09},
  year={1977},
  journaltitle = {IEEE Transactions on Computers},
  volume = {C-26},
  number = {9},
  pages = {917--922},
  issn = {1557-9956},
  doi = {10.1109/TC.1977.1674939},
  eventtitle = {{{IEEE Transactions}} on {{Computers}}}
}

\end{document}

\maketitle
{\small
$^{\text{1}}$ Life and Medical Sciences (LIMES) Institute and Bonn Center for Mathematical Life Sciences, University of Bonn, Bonn, Germany\\
$^\ast$ To whom correspondence should be addressed: jan.hasenauer@uni-bonn.de and dilan.pathirana@uni-bonn.de
}

%%%%

%\linenumbers

\section{All plausible models can be identified by relaxing the bound}
In real applications, a common issue is that data are not sufficiently informative to identify a single optimal model, i.e., the modeler might observe that multiple models describe the data sufficiently well compared to the optimal model, and that it's plausible that a different data realization will identify a different optimal model. Hence, an ensemble of plausible models needs to be constructed, e.g. by collecting all models that have a criterion value within some threshold $\Delta_\Crit$ of the optimal model. The threshold commonly boils down to some heuristic, e.g., a typical threshold choice for the AIC is $\Delta_\AIC = 10$ \citep{Burnham2004}.

This can be implemented by adjusting the condition ((2) in main text) as
\begin{equation}
    \Bound_\Crit(\Msub, \Mcand) > \Crit(\MoptCurr, \D) + \Delta_\Crit.
    \label{eq:condition_plausible}
\end{equation}
The evaluation of the condition with simpler criterion can still be skipped for submodels. For example, with the AIC, the lower bound on the number of parameters ((5) in main text) becomes
\begin{equation*}
|\Msub| > \frac{1}{2}\left(\AIC(\MoptCurr,\D)+\Delta_\AIC\right) - \frac{1}{2}\LL(\Mcand, \D).
\end{equation*}

In the remaining sections, we use the condition for optimal model(s) ((2) in main text). However, the condition for plausible models ((9) in main text) can equivalently be used -- the only difference is that a set of calibrated plausible models is collected throughout the algorithm, defined by $\{\Mcand: \Crit(\Mcand, \D) \leq \Crit(\MoptCurr, \D) + \Delta_\Crit \}$. This set needs to be re-evaluated whenever $\MoptCurr$ changes, or once after the model selection method terminates; however, these condition checks are computationally cheap if $\Crit(\Mcand)$ is saved to avoid re-computing the costly calibration step.

\section*{Bounds for specific non-monotonic criteria}

Each of the bounds derived here can be used directly in the condition for the optimal model ((2) in main text) or the condition for all plausible models \eqref{eq:condition_plausible}. For all criteria, we additionally derive an explicit bound on $|\M|$, similarly to the AIC, for faster exclusions.

\subsection*{Bayesian information criterion (BIC)}
From the definition of the BIC \citep{Schwarz1978},
\begin{align*}
    \BIC(\Msub, \D) & = 2 \NLL(\Msub, \D)  + \log (|\D|)|\Msub| \\
& \geq 2\NLL(\Mcand, \D)  + \log (|\D|)|\Msub|,
\end{align*}
hence the BIC bound is $\Bound_\BIC(\Msub, \Mcand)=2\mathrm{\NLL}(\Mcand, \D)  + \log (|\D|)|\Msub|$.
A bound on $|\Msub|$ for all submodels of $\Mcand$ that can be excluded, can be derived by substituting $\Bound_\BIC$ into the condition for the optimal model ((2) in main text),
\begin{align*}
2\NLL(\Mcand, \D)  + \log (|\D|)|\Msub| &> \BIC(\MoptCurr, \D)\\
|\Msub| &> \frac{\BIC(\MoptCurr, \D) - 2\NLL(\Mcand, \D)}{ \log (|\D|)}.
\end{align*}

\subsection*{Corrected Akaike information criterion (AIC\textsubscript{c})}

From the definition of the $\AICc$ \citep{Burnham2004},
\begin{align*}
\AICc(\Msub, \D) &= 2\NLL(\Msub, \D) + \frac{2|\Msub||\D|}{|\D| - |\Msub| - 1}\\
&\geq 2\NLL(\Mcand, \D) + \frac{2|\Msub||\D|}{|\D| - |\Msub| - 1},
\end{align*}
hence the $\AICc$ bound is $\Bound_{\AICc}(\Msub)= 2\NLL(\Mcand, \D) + \frac{2|\Msub||\D|}{|\D| - |\Msub| - 1}$.
A bound on $|\Msub|$ for all submodels of $\Mcand$ that can be excluded, can be derived by substituting $\Bound_{\AICc}$ into the condition for the optimal model ((2) in main text),
\begin{equation*}
2\NLL(\Mcand, \D) + \frac{2|\Msub||\D|}{|\D| - |\Msub| - 1} > \AICc(\MoptCurr, \D).
\end{equation*}
Re-arranging this for $|\Msub|$ involves multiplication/division by possibly-negative numbers. Hence,
\begin{equation*}
|\Msub| > \begin{cases} 
          \nBound_{\AICc} & \text{sgn}(|\D| - |\Msub| - 1) = \text{sgn}(2|\D| + \AICc(\MoptCurr, \D) - 2\NLL(\Mcand, \D)), \\
          -\nBound_{\AICc} & \text{otherwise},
       \end{cases}
\end{equation*}
where $\nBound_{\AICc}:=\frac{(\AICc(\MoptCurr, \D) - 2\NLL(\Mcand, \D))(|\D| - 1)}{2|\D|+\AICc(\MoptCurr, \D)-2\NLL(\Mcand, \D)}$ and $\text{sgn}$ is the sign function.

\subsection*{Mallows's \texorpdfstring{$\Cp$}}
Given the residual sum of squares for model $\Msub$, $\RSS(\Msub, \D)$, and the mean squared error of the superset model, $\MSE$, then, from the definition of Mallows's $\Cp$ \citep{Mallows1973},
\begin{align*}
\Cp(\Msub, \D) &= \frac{\RSS(\Msub, \D)}{\MSE} - |\D| + 2 (|\Msub| + 1),\\
 &\geq \frac{\RSS(\Mcand, \D)}{\MSE} - |\D| + 2 (|\Msub| + 1)
\end{align*}
hence the Mallows's $\Cp$ bound is $\Bound_{\Cp}(\Msub)=\frac{\RSS(\Mcand, \D)}{\MSE} - |\D| + 2 (|\Msub| + 1)$.
A bound on $|\Msub|$ for all submodels of $\Mcand$ that can be excluded, can be derived by substituting $\Bound$ into the condition for the optimal model ((2) in main text), 
\begin{align*}
\frac{\RSS(\Mcand, \D)}{\MSE} - |\D| + 2 (|\Msub| + 1) &> \Cp(\MoptCurr, \D)\\
|\Msub| &> \frac{1}{2}\left(\Cp(\MoptCurr, \D)+|\D|-\frac{\RSS(\Mcand, \D)}{\MSE}\right)-1.
\end{align*}

\textbf{Equivalence to the $\AIC$}. Assuming linear models and that the measurement error is known and given by $\sigma^2 = \MSE$, then Mallows's $\Cp$ is equivalent to the $\AIC$. Specifically, under these assumptions, $\AIC(\Mcand, \D) - \Cp(\Mcand, \D) = |\D| - 2 + \frac{1}{2} \log(2 \pi \sigma^2)$. As this difference is independent of $\Mcand$, then AIC and Mallows's $\Cp$ will only differ by a constant factor in the context of a single model selection problem, and hence will select the same optimal model(s). However, in the Hald problem, the assumption on the known variance is not met.

\subsection*{Exemplary problems}
\subsubsection*{Small-scale problem}
The model contains an ODE with states $x = (x_A, x_B)$ and initial values $x(0)=(0, 0)$. The superset model ODE system is
\begin{align}
\begin{aligned}
    \frac{dx_A}{dt} &= \theta_1 - \Big(\theta_2  + \theta_3 \cdot x_B\Big) \cdot  x_A, \\
        \frac{dx_B}{dt} &= \Big(\theta_2 - \theta_3 \cdot x_B \Big) \cdot x_A.
\end{aligned}
\end{align}
Synthetic data were generated by simulating the ground-truth model ($\theta = (0.2, 0.1, 0)$), adding additive Gaussian noise (standard deviation = 0.15), and saving the noisy $x_B$ values at time points 0, 1, 5, 10, 30, and 60. We used \texttt{AMICI} and \texttt{pyPESTO} for synthetic data generation \citep{Frohlich2021, Schaelte2023}. The full model selection problem, including the synthetic data, is available on Zenodo (\url{https://doi.org/10.5281/zenodo.14906119}).

For parameter estimation and model selection, the standard deviation is considered unknown and estimated, along with the estimated parameters in $\theta$.

Criterion values of each model were computed for each of the AIC, $\AICc$, and $\BIC$, and are displayed in Table \ref{tab:criteria_small_scale}. In principle, the different criteria will often select different models; however, in this example, the same model was selected by all criteria, and matched the ground-truth model.
\begin{table}[ht]
    \centering
    \caption{\textbf{Criterion values for all models in the small-scale ODE model selection problem}}
    \begin{tabular}{c|c||c|c|c}
    Model &    NLL &     AIC  &     $\AICc$ &    BIC \\
    \hline \hline
       $\M_{000}$ &  27.035 &  54.071  &  55.404&  59.041 \\
       \hline
       $\M_{001}$ &  27.035 &  60.071  &  59.071&  61.526 \\
       \hline
       $\M_{010}$ &  27.035 &  60.071  &  59.071&  61.526 \\
       \hline
       $\M_{011}$ &  27.035 &  62.071  &  63.785&  64.010 \\
       \hline
       $\M_{100}$ &  21.706 &  49.412 &  48.412 &  50.866 \\
       \hline
       $\M_{101}$ &  21.706 &  51.412 &  53.126 &  53.351 \\
       \hline
       $\M_{110}$ & \textbf{-12.617} & \textbf{-17.233} & \textbf{-15.519} & \textbf{-15.294} \\
       \hline
       $\M_{111}$ & --13.113 & -16.226 & -10.226 & -13.801 \\
    \end{tabular}
        \label{tab:criteria_small_scale}
\end{table}

\subsubsection*{Large-scale problem}
The ODE system, taken from \citep{Gabel2019}, is
\begin{align}
\begin{aligned}
    \frac{dx_A}{dt} &= (\rho_A - \mu_{AB} - \mu_{AC} - \mu_{AD})  \cdot x_A  + \mu_{BA} \cdot x_B + \mu_{CA} \cdot x_C + \mu_{DA} \cdot x_D,\\
    \frac{dx_B}{dt} &= \mu_{AB} \cdot x_A + (\rho_B - \mu_{BA} - \mu_{BC} - \mu_{BD})  \cdot x_B + \mu_{CB} \cdot x_C + \mu_{DB} \cdot x_D,  \\
    \frac{dx_C}{dt} &= \mu_{AC} \cdot x_A + \mu_{BC} \cdot x_B 
    + (\rho_C - \mu_{CA} - \mu_{CB} - \mu_{CD})  \cdot x_C
    + \mu_{DC} \cdot x_D,\\
    \frac{dx_D}{dt} &= \mu_{AD} \cdot x_A + \mu_{BD} \cdot x_B
    + \mu_{CD} \cdot x_C
    + (\rho_D - \mu_{DA} - \mu_{DB} - \mu_{DC})  \cdot x_D.
\end{aligned}
\end{align}
The ODE state is $x = (x_A, x_B, x_C, x_D)$, with initial state $x(0) = (100, 0, 0, 0)$. The data are observations of all states at every unit time point until 30 time units. The model used for synthetic data generation (again with \texttt{AMICI} and \texttt{pyPESTO}) had $\rho_B = \rho_C = \mu_{AB} = 0.1, \mu_{BC} = 0.05$ and $\mu_{BD} = 0.2$, and all other parameters set to zero. Gaussian noise was added to every synthetic measurement, with standard deviation = 1, which was fixed to 1 during calibration. Parameters were log-transformed for better optimizer convergence during calibration.

A summary of the performance of all methods is provided in Table \ref{tab:summary_famos}.

\begin{table}[ht]
    \centering
    \caption{\textbf{Methods applied to the Large-scale problem with AIC.}}
    \begin{tabular}{l|c|c|c|c|c}
        & forward & backward & \begin{tabular}[t]{c}\bound{}\\backward\end{tabular}  & brute-force & \begin{tabular}[t]{c}\bound{}\\exhaustive\end{tabular} \\
        \hline
        Calibrated models [\#] & 122 & 127 & 72 & 65,536 & 72 \\
        \hline
        Calibrated models [\%] & 0.19\% & 0.19\% & 0.11\% & 100.0\% & 0.11\% \\
        \hline
        Guarantees optimality & $\times$ & $\times$ & $\times$ & \checkmark & \checkmark \\
        \hline
        Identified $\Mopt$ & $\times$ & $\checkmark$ & $\checkmark$ & $\checkmark$ & $\checkmark$
    \end{tabular}
    \label{tab:summary_famos}
\end{table}

\subsubsection*{The Hald problem}
The full dataset is taken from \citep{Hald1952} (Table \ref{tab:hald_problem_data}). In addition to the covariate effect parameters $\theta$, all models additionally estimated an intercept, and residual variance, under the assumption of centered-Gaussian--distributed residuals. Model calibration was performed with \texttt{scikit-learn}.

Using AIC, $\AICc$, or BIC did not change the models that were calibrated with each method. However, using Mallows's $\Cp$ resulted in different calibrated models. Although the literature often suggests that the AIC and Mallows's $\Cp$ are equivalent for linear models with Gaussian noise, this is not the case here, because the residual variance is estimated. All models were fit to the data in Table \ref{tab:hald_problem_data}.

\begin{table}[ht]
    \centering
    \caption{\textbf{Heat of hardening of cement}. Adapted from \citep{Burnham2004}.
    The covariates of the heat of hardening are the cement ingredients, with units \% of cement clinker.}
    \begin{tabular}{c|c|c|c||c}
        \multicolumn{1}{c|}{\begin{tabular}[c]{@{}c@{}}calcium\\aluminate\\{[\%]}\end{tabular}} & 
        \multicolumn{1}{c|}{\begin{tabular}[c]{@{}c@{}}tricalcium\\silicate\\{[\%]}\end{tabular}} & 
        \multicolumn{1}{c|}{\begin{tabular}[c]{@{}c@{}}tetracalcium\\alumino\\ferrite {[\%]}\end{tabular}} & 
        \multicolumn{1}{c||}{\begin{tabular}[c]{@{}c@{}}dicalcium\\silicate\\{[\%]}\end{tabular}} &
        \begin{tabular}[c]{@{}c@{}}heat of\\ hardening\\{[$\frac{\mathrm{cal}}{\mathrm{g}}$]}\end{tabular} \\ \hline
        7 & 26 & 6 & 60 & 78.5 \\ \hline
        1 & 29 & 15 & 52 & 74.3 \\ \hline
        11 & 56 & 8 & 20 & 104.3 \\ \hline
        11 & 31 & 8 & 47 & 87.6\\ \hline
        7 & 52 & 6 & 33 & 95.9\\ \hline
        11 & 55 & 9 & 22 & 109.2\\ \hline
        3 & 71 & 17 & 6 & 102.7\\ \hline
        1 & 31 & 22 & 44 & 72.5\\ \hline
        2 & 54 & 18 & 22 & 93.1\\ \hline
        21 & 47 & 4 & 26 & 115.9\\ \hline
        1 & 40 & 23 & 34 & 83.8\\ \hline
        11 & 66 & 9 & 12 & 113.3\\ \hline
        10 & 68 & 8 & 12 & 109.4\\
        \end{tabular}
    \label{tab:hald_problem_data}
\end{table}

\begin{table}[ht]
    \centering
    \caption{\textbf{Methods applied to the Hald dataset with AIC, $\AICc$, and BIC.}}
    \begin{tabular}{l|c|c|c|c|c}
        & forward & backward & \begin{tabular}[t]{c}\bound{}\\backward\end{tabular}  & brute-force & \begin{tabular}[t]{c}\bound{}\\exhaustive\end{tabular} \\
        \hline
        Calibrated models [\#] & 10 & 8 & 6 & 16 & 6 \\
        \hline
        Calibrated models [\%] & 62.5\% & 50\% & 37.5\% & 100.0\% & 37.5\% \\
        \hline
        Guarantees optimality & $\times$ & $\times$ & $\times$ & \checkmark & \checkmark \\
        \hline
        Identified $\Mopt$ & $\checkmark$ & $\checkmark$ & $\checkmark$ & $\checkmark$ & $\checkmark$
    \end{tabular}
    \label{tab:hald_methods_ic}
\end{table}

\begin{table}[ht]
    \centering
    \caption{\textbf{Methods applied to the Hald dataset with Mallows's $\Cp$.}}
    \begin{tabular}{l|c|c|c|c|c}
        & forward & backward & \begin{tabular}[t]{c}\bound{}\\backward\end{tabular}  & brute-force & \begin{tabular}[t]{c}\bound{}\\exhaustive\end{tabular} \\
        \hline
        Calibrated models [\#] & 9 & 10 & 9 & 16 & 10 \\
        \hline
        Calibrated models [\%] & 56.25\% & 62.5\% & 56.25\% & 100.0\% & 62.5\% \\
        \hline
        Guarantees optimality & $\times$ & $\times$ & $\times$ & \checkmark & \checkmark \\
        \hline
        Identified $\Mopt$ & $\times$ & $\checkmark$ & $\checkmark$ & $\checkmark$ & $\checkmark$
    \end{tabular}
    \label{tab:hald_method_cp}
\end{table}

\subsubsection*{The Blasi et al. problem}
The Blasi et al. \citep{Blasi2016} problem involves histone H4, a protein that affects the accessibility of DNA in cell nuclei, and can be independently acetylated at four lysine sites, namely K5, K8, K12 and K16. Hence, there exist $2^4=16$ motifs (acetylation states). Each motif can be acetylated at an unacetylated site, or deacetylated at an acetylated site.

An ODE system describes the processes associated with enzymatic acetylation of histone H4, and deacetylation. These ODEs represent the instantaneous change in the abundances of the H4 motifs.

Consider some H4 motif $M$, all reactions that can acetylate another motif to produce $M$ (input reactions $I_M$), and all reactions that can acetylate $M$ into another motif (output reactions $O_M$). Deacetylation occurs with some constant rate constant $\delta$ at all sites that are currently acetylated, indexed by $A_M$. Motifs with one additional acetylated site than $M$ can be deacetylated at that site to produce $M$, and these reactions are indexed by $D_M$. Then, assuming linear kinetics, the ODE for $M$ is
\begin{equation}
\begin{split}
\frac{\mathrm{d}M}{\mathrm{d}t} = &\sum_{i \in I_M} k_i M_i - \sum_{o\in O_M} k_o M \\
& -\sum_{a\in A_M}\delta M + \sum_{d\in D_M}\delta M_d
\end{split}
\end{equation}
where $t$ is time, $k$ are reaction rate constants, and $M_i$ and $M_d$ are the corresponding motifs for those reactions.

The observations are the motifs $M(t)$ at steady-state, i.e. $d M(t) / dt = 0$. The data and parameter estimation problem are taken from the original publication \cite{Blasi2016}, and were used in the PEtab format \citep{SchmiesterSch2021}, available at the Benchmark Collection of PEtab problems \citep{BenchmarkingInitiativeBenchmarkModelsPEtabBenchmark}.

The model selection problem was also taken from the original publication \cite{Blasi2016}. Briefly, each acetylation reaction of each motif is catalyzed by an enzyme. This enzyme may catalyze all reactions in the same way with some default rate constant, or with motif-specific rate constants. In terms of the ODE system, the model selection problem is to determine which rate constants $k$ are different to the default. In total, there are 32 acetylation reactions. This yields a model space with $2^{32}=4,294,967,296$ ($>4$ billion) models.

\printbibliography